\documentclass[point_reset_at_theo]{ian}

\usepackage{amssymb}
\usepackage{graphicx}
\usepackage{largearray}
\usepackage{dsfont}
\usepackage{constants}

\begin{document}

\hbox{}
\hfil{\bf\Large
High-fugacity expansion, Lee-Yang zeros and\par
\vskip10pt
\hfil order-disorder transitions in hard-core lattice systems
}
\vskip80pt

\hfil{\bf\large Ian Jauslin}\par
\hfil{\it School of Mathematics, Institute for Advanced Study}\par
\hfil{\tt\color{blue}\href{mailto:jauslin@ias.edu}{jauslin@ias.edu}}
\vskip20pt

\hfil{\bf\large Joel L. Lebowitz}\par
\hfil{\it Departments of Mathematics and Physics, Rutgers University}\par
\hfil{\it Simons Center for Systems Biology, Institute for Advanced Study}\par
\hfil {\tt\color{blue}\href{mailto:lebowitz@math.rutgers.edu}{lebowitz@math.rutgers.edu}}

\vskip80pt

\hfil {\bf Abstract}\par
\medskip
We establish existence of order-disorder phase transitions for a class of ``non-sliding'' hard-core lattice particle systems on a lattice in two or more dimensions. All particles have the same shape and can be made to cover the lattice perfectly in a finite number of ways. We also show that the pressure and correlation functions have a convergent expansion in powers of the inverse of the fugacity. This implies that the Lee-Yang zeros lie in an annulus with finite positive radii.

\vfill

\tableofcontents

\eject

\setcounter{page}1
\pagestyle{plain}

\section{Introduction}
\indent One of the most important open problems in the theory of equilibrium statistical mechanics, is to prove the existence of order-disorder phase transitions in continuum particle systems. While such fluid-crystal transitions are ubiquitous in real systems and are observed in computer simulations of systems with  effective pair potentials, there are no proofs, or even good heuristics, for showing this mathematically. A paradigmatic example of this phenomenon is the fluid-crystal transition for hard spheres in 3 dimensions, observed in simulations and experiments~\-\cite{WJ57,AW57,PM86,IK15}. Whereas, in 2 dimensions, crystalline states are ruled out by the Mermin-Wagner theorem~\-\cite{Ri07}, it is believed that there are other transitions for hard discs~\-\cite{BK11} (see~\-\cite{St88} or~\-\cite[section~8.2.3]{Mc10} for a review), though none have, as of yet, been proven. Such transitions are purely geometric. They are driven by entropy and depend only on the density, that is, on the volume fraction taken up by the hard particles.
\bigskip

\begin{figure}
  \hfil\includegraphics[width=2cm]{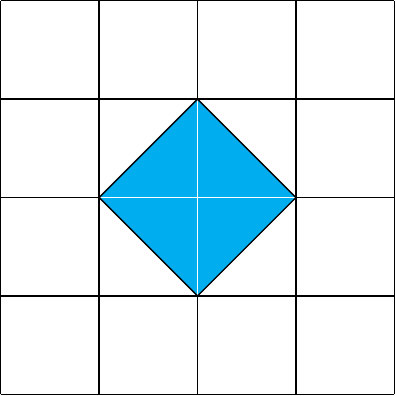} {\footnotesize\it a.}
  \hfil\includegraphics[width=2cm]{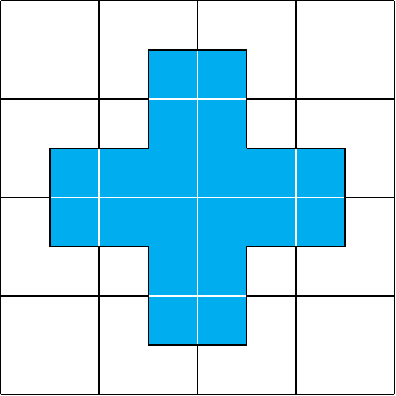} {\footnotesize\it b.}
  \hfil\includegraphics[width=2cm]{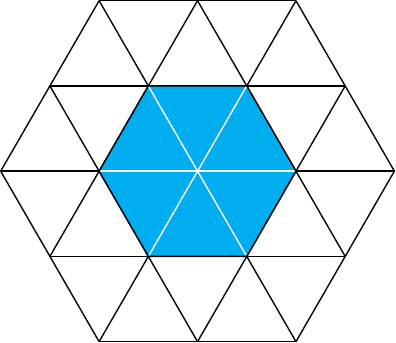} {\footnotesize\it c.}
  \caption{Three non-sliding hard-core lattice particle systems.\par
     {\it a}. The hard diamond model is equivalent to the nearest neighbor exclusion on $\mathbb Z^2$.\par
     {\it b}. The hard cross model is equivalent to the third-nearest neighbor exclusion on $\mathbb Z^2$.\par
     {\it c}. The hard hexagon model is equivalent to the nearest neighbor exclusion on the triangular lattice.
  }
  \label{fig:shapes}
\end{figure}

\indent The situation is different for lattice systems, where there are many examples for which such entropy-driven transitions have been proven. A simple example is that of hard ``diamonds'' on the square lattice (see figure~\-\ref{fig:shapes}{\it a}), which is a model on $\mathbb Z^2$ with nearest-neighbor exclusion. As was shown by Dobrushin~\-\cite{Do68}, this model transitions from a low-density disordered state to a high-density crystalline phase, where the even or odd sublattice is preferentially occupied. The heuristics of this transition had been understood earlier (the hard diamond model is related to the 0-temperature limit of the antiferromagnetic Ising model for which the exponential of the magnetic field plays the role of the fugacity~\-\cite{BK73,LRS12}), for instance by Gaunt and Fisher~\-\cite{GF65}, who extrapolated a low- and high-fugacity expansion of the pressure $p(z)$ to find a singularity at a critical fugacity $z_t>0$. A similar analysis was carried out for the nearest neighbor exclusion on $\mathbb Z^3$ by Gaunt~\-\cite{Ga67}.

\indent The low-fugacity expansion in powers of the fugacity $z$ dates back to Ursell~\-\cite{Ur27} and Mayer~\-\cite{Ma37}. Its radius of convergence was bounded below by Groeneveld~\-\cite{Gr62} for positive pair-potentials and by Ruelle~\-\cite{Ru63} and Penrose~\-\cite{Pe63} for general pair-potentials.

\indent The high-fugacity expansion is an expansion in powers of the inverse fugacity $y\equiv z^{-1}$. As far as we know, it was first considered by Gaunt and Fisher~\-\cite{GF65} for the hard diamond model, without any indication of its having a positive radius of convergence, or that its coefficients are finite in the thermodynamic limit beyond the first 9 terms.
\bigskip

\indent In this paper we prove, using an extension of Pirogov-Sinai theory~\-\cite{PS75,KP84}, that the high-fugacity expansion has a positive radius of convergence for a class of hard-core lattice particle systems in $d\geqslant 2$ dimensions. We call these {\it non-sliding} models. In addition, we show that these systems exhibit high-density crystalline phases, which, combined with the convergence of the low-fugacity expansion proved in~\-\cite{Gr62,Ru63,Pe63}, proves the existence of an order-disorder phase transition for these models. A preliminary account of this work, without proofs, is in~\-\cite{JL17}.
\bigskip

\indent {\it Non-sliding} models are systems of identical hard particles which have a finite number $\tau$ of maximal density perfect coverings of the infinite lattice, and are such that any defect in a covering (a defect appears where a particle configuration differs from a perfect covering) leaves an amount of empty space that is proportional to its size, and that a particle configuration is characterized by its defects (this will be made precise in the following). This class includes all of the models for which crystallization has been proved, like the hard diamond~\-\cite{Do68} (see figure~\-\ref{fig:shapes}{\it a}) model discussed above, as well as the hard cross model~\-\cite{HP74} (see figure~\-\ref{fig:shapes}{\it b}), which corresponds to the third-nearest-neighbor exclusion on $\mathbb Z^2$, and the hard hexagon model on the triangular lattice~\-\cite{Ba82} (see figure~\-\ref{fig:shapes}{\it c}), which corresponds to the nearest-neighbor exclusion on the triangular lattice.

\indent The hard diamond model was studied by Gaunt and Fisher~\-\cite{GF65}, in which the first 13 terms of the low-fugacity expansion and the first 9 terms of the high-fugacity expansion were computed, from which Gaunt and Fisher predicted a phase transition at the point where both expansions, suitably extrapolated, meet.

\indent The hard cross model was studied by Heilmann and Pr\ae stgaard~\-\cite{HP74}, who gave a sketch of a proof that it has a crystalline high-density phase. Eisenberg and Baram~\-\cite{EB05} computed the first 13 terms of the low-fugacity expansion and the first 6 terms of the high-fugacity expansion for this model, and conjectured that it should have a {\it first-order} order-disorder phase transition. We will prove the convergence of the high-fugacity expansion, and reproduce Heilmann and Pr\ae stgaard's result, but will stop short of proving the order of the phase transition, for which new techniques would need to be developed. We will also extend this result to the hard cross model on a fine lattice, although the present techniques do not allow us to go to the continuum.

\indent The hard hexagon model on the triangular lattice was shown to be exactly solvable by Baxter~\-\cite{Ba80,Ba82}, and to be crystalline at high densities. The exact solution provides an (implicit) expression for the pressure $p(z)$, from which the high-fugacity expansion can be obtained, as shown by Joyce~\-\cite{Jo88}.
\bigskip

\subsection{Hard-core lattice particle models}\label{sec:model}
\indent Consider a $d$-dimensional lattice $\Lambda_\infty$. We consider $\Lambda_\infty$ as a graph, that is, every vertex of $\Lambda_\infty$ is assigned a set of {\it neighbors}. We denote the graph distance on $\Lambda_\infty$ by $\Delta$, in terms of which $x,x'\in\Lambda_\infty$ are neighbors if and only if $\Delta(x,x')=1$. We will consider systems of identical particles on $\Lambda_\infty$ with hard core interactions. We will represent the latter by assigning a {\it support} to each particle, which is a connected and bounded subset $\omega\subset\mathbb R^d$ (we need not assume much about $\omega$, because we will mainly consider its intersections with the lattice), and forbid the supports of different particles from intersecting. In the examples mentioned above, the shapes would be a diamond, a cross or a hexagon (see figure~\-\ref{fig:shapes}). Note that $\omega$ may, in some cases be an open set, whereas in others, it might include a portion of its boundary (see section~\-\ref{section:non_sliding} for details). We define the grand-canonical partition function of the system at activity $z>0$ on any bounded $\Lambda\subset\Lambda_\infty$ as
\begin{equation}
  \Xi_\Lambda(z)=\sum_{X\subset\Lambda}z^{|X|}\prod_{x\neq x'\in X}\varphi(x,x')
  \label{Xi}
\end{equation}
in which $X$ is a particle configuration in $\Lambda$ (that is, a set of lattice points $x\in\Lambda$ on which particles are placed), $|X|$ is the cardinality of $X$, and, denoting $\omega_x:= \{x+y,\ y\in\omega\}$ ($\omega_x$ is the support of the particle {\it located} at $x$), $\varphi(x,x')\in\{0,1\}$ enforces the hard core repulsion: it is equal to 1 if and only if $\omega_{x}\cap\omega_{x'}=\emptyset$. In the following, a subset $X\subset\Lambda_\infty$ is said to be a {\it particle configuration} if $\varphi(x,x')=1$ for every $x\neq x'\in X$, and we denote the set of particle configurations in $\Lambda$ by $\Omega(\Lambda)$. We define $N_{\mathrm{max}}$ as the maximal number of particles:
\begin{equation}
  N_{\mathrm{max}}:=\mathrm{max}\{|X|,\ X\subset\Lambda\}
  .
\end{equation}
In addition, note that several different shapes can, in some cases, give rise to the same partition function. For example, the hard diamond model is equivalent to a system of hard disks of radius $r$ with $\frac12<r<\frac1{\sqrt 2}$.
\bigskip

\indent We will discuss the properties of the finite-volume {\it pressure} of hard-core particles systems, defined as
\begin{equation}
p_\Lambda(z):=\frac1{|\Lambda|}\log \Xi_\Lambda(z)
\label{p}
\end{equation}
and its infinite-volume limit
\begin{equation}
  p(z):=\lim_{\Lambda\to\Lambda_\infty}p_\Lambda(z)=:\rho_m\log z+f(y)
\end{equation}
in which $y\equiv z^{-1}$ and $\rho_m$ is the maximal density $\rho_m=\lim_{\Lambda\to\Lambda_\infty}\frac{N_{\mathrm{max}}}{|\Lambda|}$. In particular, we will focus on the analyticity properties of $f(y)$. When $f(y)$ is analytic for small values of $y$, the system is said to admit a convergent {\it high-fugacity} expansion.

\subsection{Low-fugacity expansion}\label{sec:low_fugacity}
\indent The main ideas underlying the high-fugacity expansion come from the low-fugacity expansion, which we will now briefly review. It is an expansion of $p_\Lambda$ in powers of the fugacity $z$, and its formal derivation is fairly straightforward: defining the {\it canonical} partition function as
\begin{equation}
  Z_\Lambda(k):=\sum_{\displaystyle\mathop{\scriptstyle X\subset\Lambda}_{|X|=k}}\prod_{x\neq x'\in X}\varphi(x,x')
\end{equation}
as the number of particle configurations with $k$ particles, (\ref{Xi}) can be rewritten as
\begin{equation}
  \Xi_\Lambda(z)=\sum_{k=0}^{N_{\mathrm{max}}} z^kZ_\Lambda(k).
  \label{Xi_z}
\end{equation}
Injecting~\-(\ref{Xi_z}) into~\-(\ref{p}), we find that, formally,
\begin{equation}
  p_\Lambda(z)=\sum_{k=1}^\infty b_k(\Lambda)z^k
  \label{p_z}
\end{equation}
with
\begin{equation}
  b_k(\Lambda):=\frac1{|\Lambda|}\sum_{n=1}^k\frac{(-1)^{n+1}}n\sum_{\displaystyle\mathop{\scriptstyle k_1,\cdots,k_n\geqslant 1}_{k_1+\cdots+k_n=k}}Z_\Lambda(k_1)\cdots Z_\Lambda(k_n).
  \label{blog}
\end{equation}
As was shown in~\-\cite{Ur27,Ma37,Gr62,Ru63,Pe63}, there is a remarkable cancellation that eliminates the terms in $b_k(\Lambda)$ that diverge as $\Lambda\to\Lambda_\infty$, so that $b_k(\Lambda)\to b_k$ when $\Lambda\to\Lambda_\infty$. This becomes obvious when the $b_k(\Lambda)$ are expressed as integrals over Mayer graphs. In addition, the radius of convergence $R(\Lambda)$ of~\-(\ref{p_z}) converges to $R>0$, which is at least as large as the radius of convergence of $\sum_{k=1}^\infty b_kz^k$ (for positive pair potentials, $R$ is {\it equal} to the radius of convergence~\-\cite{Pe63}).

\subsection{High-fugacity expansion}
\indent The low-fugacity expansion is obtained by perturbing around the vacuum state by adding particles to it. The high-fugacity expansion will be obtained by perturbing perfect coverings by introducing {\it defects}. Single-particle defects, corresponding to removing one particle from a perfect covering, come with a cost $y\equiv z^{-1}$, which is, effectively, the fugacity of a hole. The main idea, due to Gaunt and Fisher~\-\cite{GF65}, is to carry out a cluster expansion for the defects, which is similar to the low-fugacity expansion described above. Let us go into some more detail in the example of the hard diamond model.
\bigskip

\indent We will take $\Lambda$ to be a $2n\times 2n$ torus, which can be completely packed with diamonds (see figure~\-\ref{fig:diamond_packing}). The number of perfect covering configurations is
\begin{equation}
  \tau=2
\end{equation}
and the maximal number of particles and maximal density are
\begin{equation}
  N_{\mathrm{max}}=\rho_m|\Lambda|
  ,\quad
  \rho_m=\frac12.
\end{equation}
We denote the number of configurations that are {\it missing} $k$ particles as
\begin{equation}
  Q_\Lambda(k):=Z_\Lambda(N_{\mathrm{max}}-k)
\end{equation}
in terms of which
\begin{equation}
  \Xi_\Lambda(z)=\tau z^{N_{\mathrm{max}}}\sum_{k=0}^{N_{\mathrm{max}}} \left(\frac1\tau z^{-k}Q_\Lambda(k)\right)
\end{equation}
(we factor $\tau$ out because $Q_\Lambda(0)=\tau$ and we wish to expand the logarithm in~\-(\ref{p}) around 1). We thus have, formally
\begin{equation}
  p_\Lambda(y)=\frac1{|\Lambda|}\log\tau+\rho_m\log z+\sum_{k=1}^{N_{\mathrm{max}}} c_k(\Lambda)y^k
  \label{p_y}
\end{equation}
where $y\equiv z^{-1}$ and
\begin{equation}
  c_k(\Lambda):=\frac1{|\Lambda|}\sum_{n=1}^k\frac{(-1)^{n+1}}{n\tau^n}\sum_{\displaystyle\mathop{\scriptstyle k_1,\cdots,k_n\geqslant 1}_{k_1+\cdots+k_n=k}}Q_\Lambda(k_1)\cdots Q_\Lambda(k_n).
  \label{blog}
\end{equation}
The first 9 $c_k(\Lambda)$ are reported in~\-\cite[table~\-XIII]{GF65} and, as for the low-fugacity expansion, there is a remarkable cancellation that ensures that these coefficients converge to a finite value $c_k$ as $\Lambda\to\Lambda_\infty$. {\it But}, unlike the low-fugacity expansion, there is no {\it systematic} way of exhibiting this cancellation for general hard-core lattice particle systems. In fact there are many example of systems in which the coefficients $c_k(\Lambda)$ diverge as $\Lambda\to\Lambda_\infty$, like the nearest-neighbor exclusion model in 1 dimension (which maps, exactly, to the 1-dimensional monomer-dimer model), for which
\begin{equation}
  Q_\Lambda(1)=\frac14|\Lambda|^2
  ,\quad
  Q_\Lambda(2)=\frac1{192}(|\Lambda|^2-4)|\Lambda|^2
  ,\quad
  c_1(\Lambda)=\frac1{8}|\Lambda|
  ,\quad
  c_2(\Lambda)=-\frac1{192}|\Lambda|(5|\Lambda|^2+4).
  \label{monomer_dimer_div}
\end{equation}
Note that the pressure for this system, given by
\begin{equation}
  p(y)-\rho_m\log z=\log\left(\frac{1+\sqrt{1+4z}}2\right)-\frac12\log z
  =\log\left(\sqrt{1+\frac14y}+\frac12\sqrt y\right)
\end{equation}
is not an analytic function of $y\equiv z^{-1}$ at $y=0$ (though it is an analytic function of $\sqrt y$). Clearly, this model does not satisfy the non-sliding property. There are examples in higher dimensions of sliding models for which the pressure is not analytic in $y$, and which are not crystalline at high fugacities (see, for example, \cite{GD07}).
\bigskip

\indent One of our goals, in this paper, is to prove that, for non-sliding models, the pressure is analytic in a disk around $y=0$, thus proving the validity of the Gaunt-Fisher expansion for non-sliding systems.
\bigskip

{\bf Remark}: Let us note that, at finite temperature, lattice gases of particles with a {\it bounded} pair potential $\varphi$ that admit a convergent low-fugacity expansion (for example for summable potentials) also admit a high-fugacity expansion. This follows immediately from the spin-flip symmetry of the corresponding Ising model, which implies that
\begin{equation}
  p_\Lambda(z)=
  \log(ze^{-\frac12\alpha})
  p(ye^\alpha)
  ,\quad
  e^\alpha:=e^{\beta\sum_{x\in\Lambda}\varphi(|x|)}
  \label{particle_hole}
\end{equation}
The radius of convergence $\tilde R(\Lambda)$ of the expansion in $y$ is therefore related to the radius $R(\Lambda)$ of convergence of the expansion in $z$: $\tilde R(\Lambda)=R(\Lambda)e^{-\alpha}$. This coincides, at sufficiently high temperature, with the results of Gallavotti, Miracle-Sole and Robinson~\-\cite{GMR67}, who prove analyticity for small values of $\frac z{1+z}$. (A similar result holds for bounded many-particle interactions.)

\subsection{High-fugacity expansion and Lee-Yang zeros}
\indent As was pointed out by Lee and Yang~\-\cite{YL52,LY52}, a powerful tool to study the equilibrium properties of a system is via the positions of the roots of the partition function as a function of the fugacity $z$, called the {\it Lee-Yang zeros} of the system. In particular, the logarithm of the partition function and, consequently, the pressure, diverge at the Lee-Yang zeros, so whenever the limiting density of the roots approaches the positive real axis, this signals the presence of a phase transition. Let us denote the set of Lee-Yang zeros of a hard-core lattice particle system by $\{\xi_1(\Lambda),\cdots,\xi_{N_{\mathrm{max}}}(\Lambda)\}$. The convergence of the low-fugacity expansion within its radius of convergence $R(\Lambda)>0$ implies that every Lee-Yang zero satisfies $|\xi_i(\Lambda)|\geqslant R(\Lambda)$, and that this inequality is sharp. Similarly, when the high-fugacity expansion has a positive radius of convergence $\tilde R(\Lambda)>0$, every Lee-Yang zero must satisfy
\begin{equation}
  R(\Lambda)\leqslant |\xi_i(\Lambda)|\leqslant \tilde R(\Lambda)^{-1}
\end{equation}
and these inequalities are sharp. In addition, writing the partition function as
\begin{equation}
  \Xi_\Lambda(z)
  =\prod_{i=1}^{N_{\mathrm{max}}}\left(1-\frac z{\xi_i(\Lambda)}\right)
  =\frac{z^{N_{\mathrm{max}}}}{\prod_{i=1}^{N_{\mathrm{max}}}(-\xi_i(\Lambda))}\prod_{i=1}^{N_{\mathrm{max}}}(1-y\xi_i(\Lambda))
\end{equation}
we rewrite the high-fugacity expansion~\-(\ref{p_y}) as
\begin{equation}
  p_\Lambda(y)=\rho_m\log z-\frac1{|\Lambda|}\sum_{i=1}^{N_{\mathrm{max}}}\log(-\xi_i(\Lambda))
  -\sum_{k=1}^\infty \frac{y^k}k\left(\frac1{|\Lambda|}\sum_{i=1}^{N_{\mathrm{max}}}\xi_i^k(\Lambda)\right)
\end{equation}
which, in particular, implies that
\begin{equation}
  \prod_{i=1}^{N_{\mathrm{max}}}(-\xi_i(\Lambda))=\frac1{Q_\Lambda(0)}
  ,\quad
  c_k(\Lambda)=-\frac1k\left(\frac1{|\Lambda|}\sum_{i=1}^{N_{\mathrm{max}}}\xi_i^k(\Lambda)\right)
  .
\end{equation}
When taking the thermodynamic limit, $kc_k$ is proportional to the average of the $k$-th power of $\xi$ weighted against the limiting distribution of Lee-Yang zeros. Thus, the high-fugacity expansion converges if and only if the average of $\xi^k$ grows at most exponentially in $k$.
\bigskip

{\bf Remark}: As noted earlier, for bounded potentials, we find that the Lee-Yang zeros all lie in an annulus of radii $R(\Lambda)$ and $e^\alpha/R(\Lambda)$. Note that if one were to consider a hard-core model as the limit of a bounded repulsive potential, the hard-core limit would correspond to taking $\alpha\to\infty$. This implies that some zeros move out to infinity and that the radius of convergence of the high-fugacity expansion tends to 0 as $\alpha\to\infty$. This does not, however, imply that in the hard-core limit $\Xi_\Lambda(y)$ vanishes for $y=0$: indeed the distribution of Lee-Yang zeros does not approach the hard-core limit continuously, as is made obvious by the fact that the number of Lee-Yang zeros for finite potentials is $|\Lambda|$, whereas it is $N_{\mathrm{max}}$ in the hard-core limit. Instead, when a hard-core particle system has a convergent high-fugacity expansion, there is a bound on the remaining zeros which remains finite as $\Lambda\to\Lambda_\infty$.

\subsection{Definitions and results}
\indent We focus on the class of hard-core lattice particle models that satisfy the {\it non-sliding} property, which, roughly, means that the system admits only a finite number of perfect coverings, that any defect in a covering induces an amount of empty space that is proportional to its volume, and that any particle configuration is entirely determined by its defects. More precisely, defining $\sigma_x$ as the set of lattice sites that are covered by a particle located at $x$:
\begin{equation}
  \sigma_x:=\omega_x\cap\Lambda_\infty
  \label{sigma}
\end{equation}
given a particle configuration $X\in\Omega(\Lambda)$, we define the set of {\it empty} sites as those that are not covered by any particle:
\begin{equation}
  \mathcal E_\Lambda(X):=\{y\in\Lambda,\quad \forall x\in X,\ y\not\in\sigma_x\}
  \label{mcE}
\end{equation}
A {\it perfect covering} is defined as a particle configuration $X\in\Omega(\Lambda_\infty)$ that leaves no empty sites: $\mathcal E_{\Lambda_\infty}(X)=\emptyset$.
\bigskip

\theoname{Definition}{sliding}\label{def:sliding}
  A hard-core lattice particle system is said to be {\it non-sliding} if the following hold.
  \begin{itemize}
    \item There exists $\tau>1$, a {\it periodic} perfect covering $\mathcal L_1$, and a finite family $(f_2,\cdots,f_\tau)$ of isometries of $\Lambda_\infty$ such that, for every $i$, $\mathcal L_i\equiv f_i(\mathcal L_1)$ is a perfect covering (see figure~\-\ref{fig:cross_packing} for an example). (Here, when we use the word `lattice', we do not intend a discrete subgroup of $\mathbb R^d$ but a discrete periodic subset of $\mathbb R^d$; the sets $\mathcal L_i$ will be called `sublattices' in the following, even though they may not have any group structure.)
    \item Given a bounded {\it connected} particle configuration $X\in\Omega(\Lambda_\infty)$ (that is, a configuration in which the union $\bigcup_{x\in X}\sigma_x$ is connected), we define $\mathbb S(X)$, roughly (see~\-(\ref{bbS}) for a formal definition), as the set of particle configurations $X'$ that
    \begin{itemize}
      \item contain $X$,
      \item are such that every $x'\in X'\setminus X$ is adjacent to $X$,
      \item leave no empty sites adjacent to $\bigcup_{x\in X}\sigma_x$.
    \end{itemize}
    (see figures~\-\ref{fig:cross_unique1} and~\-\ref{fig:cross_unique2}):
    \begin{equation}
      \mathbb S(X):=
      \{X'\in\Omega(\Lambda_\infty),\ X'\supset X,\ \Delta(\mathcal E_{\Lambda_\infty}(X'),{\textstyle\bigcup_{x\in X}\sigma_x})> 1,\ \forall x'\in X', \Delta(\sigma_{x'},{\textstyle\bigcup_{x\in X}\sigma_x})\leqslant 1\}
      \label{bbS}
    \end{equation}
    in which, we recall, $\Delta$ denotes the graph distance on $\Lambda_\infty$. In order to be non-sliding, a model must be such that, for every bounded connected $X$, $\mathbb S(X)=\emptyset$, or, $\forall X'\in\mathbb S(X)$, there exists a unique $\mu\in\{1,\cdots,\tau\}$ such that $X'\subset\mathcal L_\mu$.
  \nopagebreakafteritemize
  \end{itemize}
\restorepagebreakafteritemize
\endtheo
\bigskip

{\bf Remark}: In non-sliding models, every defect (recall that a defect appears where a configuration differs from a perfect covering) induces an amount of empty space proportional to its size because any connected particle configuration $X$ that is not a subset of any perfect covering must have $\mathbb S(X)=\emptyset$, which means that there must be some empty space next to it. In addition, a particle configuration is determined by the empty space and the particles surrounding it, since the remainder of the particle configuration consists of disconnected groups, each of which is the subset of a perfect covering. The position of the particles surrounding it uniquely determines which one of the perfect coverings it is a subset of.
\bigskip

\indent In addition, we make the following assumption about the geometry of $\Lambda$: $\Lambda$ is {\it bounded}, connected and $\Lambda_\infty\setminus\Lambda$ is connected, and {\it tiled}, by which we mean that there must exist $\mu\in\{1,\cdots,\tau\}$ and a set $S\subset\mathcal L_\mu$ such that
\begin{equation}
  \Lambda=\bigcup_{x\in S}\sigma_x
  .
  \label{tiled}
\end{equation}
The choice of $\mu$ will not play any role in the thermodynamic limit.
\bigskip

\indent Given such a $\Lambda$, we will consider the following boundary conditions. Given $\nu\in\{1,\cdots,\tau\}$ (which is not necessarily equal to the $\mu$ with which we tiled $\Lambda$), we define $\Omega_\nu(\Lambda)$ as the set of particle configurations such that, roughly (see~\-(\ref{Omeganu}) for a formal definition),
\begin{itemize}
  \item every site $x\in\mathcal L_\nu$ such that $\Delta(\sigma_x,\Lambda_\infty\setminus\Lambda)\leqslant 1$, is occupied by a particle,
  \item the particles that neighbor the boundary must not neighbor an empty site in $\Lambda_\infty$.
\end{itemize}
Thus, defining $\mathbb B_\nu(\Lambda):=\{x\in\mathcal L_\nu\cap\Lambda,\ \Delta(\sigma_x,\Lambda_\infty\setminus\Lambda)\leqslant 1\}$ as the set of sites in $\mathcal L_\nu$ that neighbor the boundary, and $\mathbb X_\nu(\Lambda):=\mathcal L_\nu\setminus\Lambda$, we define
\begin{equation}
  \Omega_\nu(\Lambda):=
  \{
    X\subset\Lambda,\quad
    X\supset\mathbb B_\nu(\Lambda)
    ,\quad
    \forall x\in\mathbb B_\nu(\Lambda),\ \Delta(\sigma_x,\mathcal E_{\Lambda_\infty}(X\cup\mathbb X_\nu(\Lambda)))>1
  \}
  .
  \label{Omeganu}
\end{equation}
We choose these particular boundary conditions in order to make the discussion below simpler. Certain types of more general boundary conditions would presumably lead to infinite volume measures which are convex combinations of those induced by the boundary conditions considered here. For example, for the hard diamond model with periodic or open boundary conditions, we would get a limiting state which is a $\frac12$-$\frac12$ superposition of the even and odd states.
\bigskip

\indent Allowing the fugacity to depend on the position of the particle, we define the partition function with fugacity $\underline z:\Lambda_\infty\to[0,\infty)$ and boundary condition $\nu$ as
\begin{equation}
  \Xi_\Lambda^{(\nu)}(\underline z)=\sum_{X\in\Omega_\nu(\Lambda)}\left(\prod_{x\in X}\underline z(x)\right)\prod_{x\neq x'\in X}\varphi(x,x')
  .
  \label{Xi_nu}
\end{equation}
Since the infinite-volume pressure is independent of the boundary condition, it can be recovered from $\Xi_\Lambda^{(\nu)}(\underline z)$ by setting $\underline z(x)\equiv z$. By allowing the fugacity to depend on the position of the particle, we can compute the {\it $\mathfrak n$-point truncated correlation functions} of the system with $\nu$-boundary conditions at fugacity $z$, defined as
\begin{equation}
  \rho_{n,\Lambda}^{(\nu)}(\mathfrak x_1,\cdots,\mathfrak x_n):=
  \left.\frac{\partial^{\mathfrak n}}{\partial\log\underline z(\mathfrak x_1)\cdots\partial\log\underline z(\mathfrak x_{\mathfrak n})}
    \log\Xi_\Lambda^{(\nu)}(\underline z)
  \right|_{\underline z(x)\equiv z}
\end{equation}
as well as its infinite-volume limit
\begin{equation}
  \rho_n^{(\nu)}(\mathfrak x_1,\cdots,\mathfrak x_n)
  :=\lim_{\Lambda\to\Lambda_\infty}
  \rho_{n,\Lambda}^{(\nu)}(\mathfrak x_1,\cdots,\mathfrak x_n)
  .
\end{equation}
Note that the 1-point correlation function is the local density. In addition, we define the {\it average density} as
  \begin{equation}
    \rho:=\lim_{\Lambda\to\Lambda_\infty}\frac1{|\Lambda|}\sum_{x\in\Lambda}\rho_{1,\Lambda}^{(\nu)}(x)
    .
  \end{equation}
\bigskip

\indent Our main result is summarized in the following theorem.
\bigskip

\theoname{Theorem}{crystallization and high-fugacity expansion}\label{theo:main}
  Consider a non-sliding hard-core lattice particle system. There exists $y_0>0$ such that, if $|y|<y_0$, then there are $\tau$ distinct extremal Gibbs states. The $\nu$-th Gibbs state, obtained from the boundary condition labeled by $\nu$, is invariant under the translations of the sublattice $\mathcal L_\nu$. In addition, for any boundary condition $\nu\in\{1,\cdots,\tau\}$, any $\mathfrak n\geqslant 1$ and $\mathfrak x_1,\cdots,\mathfrak x_{\mathfrak n}\in\Lambda_\infty$, both $p(z)-\rho_m\log z$ and the $n$-point truncated correlation function $\rho_n^{(\nu)}(\mathfrak x_1,\cdots,\mathfrak x_n)$ are analytic functions of $y$ for $|y|<y_0$.
  \bigskip

  These Gibbs states are {\it crystalline}: having picked the boundary condition $\nu$, the particles are much more likely to be on the $\mathcal L_\nu$ sublattice than the others: for every $x\in\Lambda_\infty$,
  \begin{equation}
    \rho_1^{(\nu)}(x)=
    \left\{\begin{array}{ll}
      1+O(y)&\mathrm{if\ }x\in\mathcal L_\nu\\[0.3cm]
      O(y)&\mathrm{if\ not}
      .
    \end{array}\right.
    \label{crystallization}
  \end{equation}
  \bigskip

  Finally, both $p+\rho_m\log(\rho_m-\rho)$ and $\rho_n^{(\nu)}(\mathfrak x_1,\cdots,\mathfrak x_n)$ are analytic functions of $\rho_m-\rho$, with a positive radius of convergence.
\endtheo
\bigskip

{\bf Remark}: We show that the analyticity of the pressure in $y$ implies analyticity in $\rho_m-\rho$. The converse is not necessarily true. In particular, if $p-\rho_m\log z$ is analytic in $y^\alpha$ for some $\alpha$ (as is the case for the 1-dimensional nearest neighbor exclusion, for which $\alpha=\frac12$), then it is also analytic in $\rho_m-\rho$.

\section{Non-sliding hard-core lattice particle models}\label{section:non_sliding}
\indent In this section, we present several examples of non-sliding hard-core lattice particle models.
\bigskip

\begin{figure}
  \hfil\includegraphics[width=4cm]{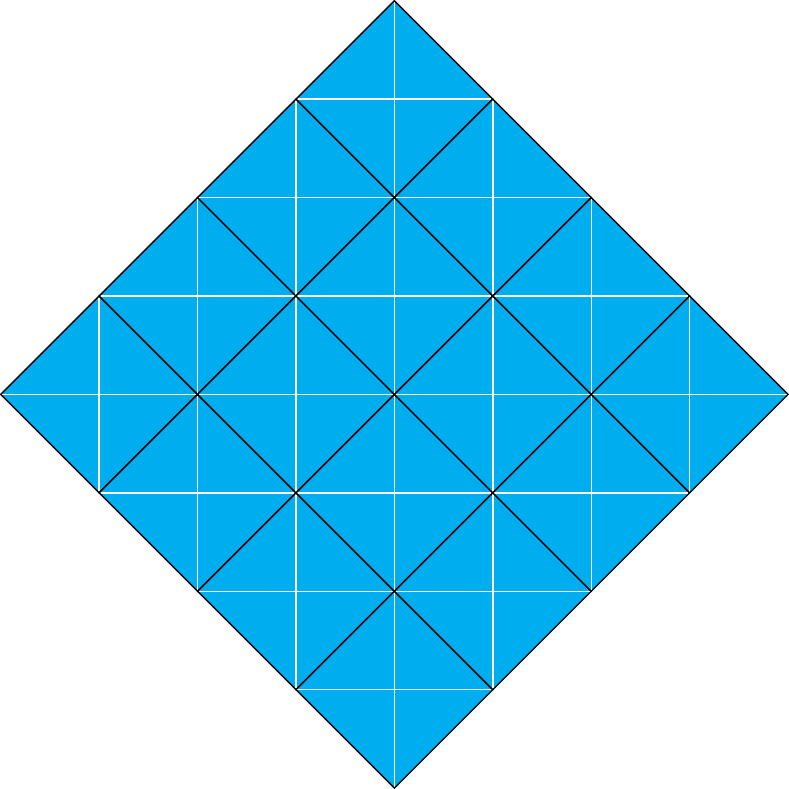}
  \caption{Perfect covering of diamonds. There are 2 inequivalent such coverings, obtained by translating the one depicted here.}
  \label{fig:diamond_packing}
\end{figure}

\point Let us start with the hard diamond model, or rather, a generalization to the ``hyperdiamond'' model in $d\geqslant 2$-dimensions, which is equivalent to the nearest neighbor exclusion on $\mathbb Z^d$. It is formally defined by specifying the lattice $\Lambda_\infty=\mathbb Z^d$ and the hyperdiamond shape $\omega\subset\mathbb R^d$ (see figure~\-\ref{fig:shapes}{\it a}):
\begin{equation}
  \omega=\left\{(x_1,\cdots,x_d)\in(-1,1)^d,\ {\textstyle\sum_{i=1}^n}|x_i|<1\}\cup\{(0,\cdots,0,1)\right\}
  .
\end{equation}
Note the adjunction of the point $(0,\cdots,0,1)$, whose absence would prevent the existence of any perfect covering (see figure~\-\ref{fig:diamond_packing}), and implies that each hyperdiamond covers two sites. The notion of {\it connectedness} in $\Lambda_\infty$ is defined as follows: two points are connected if and only if they are at distance 1 from each other. There are 2 perfect coverings (see figure~\-\ref{fig:diamond_packing}):
\begin{equation}
  \mathcal L_1=\{(x_1,\cdots,x_d)\in\mathbb Z^d,\ x_1+\cdots+x_d\mathrm{\ even}\}
  ,\quad
  \mathcal L_2=\{(x_1,\cdots,x_d)\in\mathbb Z^d,\ x_1+\cdots+x_d\mathrm{\ odd}\}
  \label{coverings_diamonds}
\end{equation}
which are related to each other by the translation by $(0,\cdots,0,1)$. Finally, this model satisfies the non-sliding condition because any pair $x_1,x_2\in\mathbb Z^d$ of hyperdiamonds whose supports are disjoint and connected (connected, here, refers to the set $\sigma_{x_1}\cup\sigma_{x_2}$) are both in the same sublattice: $(x_1,x_2)\in\mathcal L_1^2\cup\mathcal L_2^2$, and the distinct sublattices do not overlap $\mathcal L_1\cap\mathcal L_2=\emptyset$. Connected hyperdiamond configurations are, therefore, always subsets of $\mathcal L_1$ or of $\mathcal L_2$, and one can find which one it is from the position of a single one of its particles.
\bigskip

\begin{figure}
  \hfil\includegraphics[width=6cm]{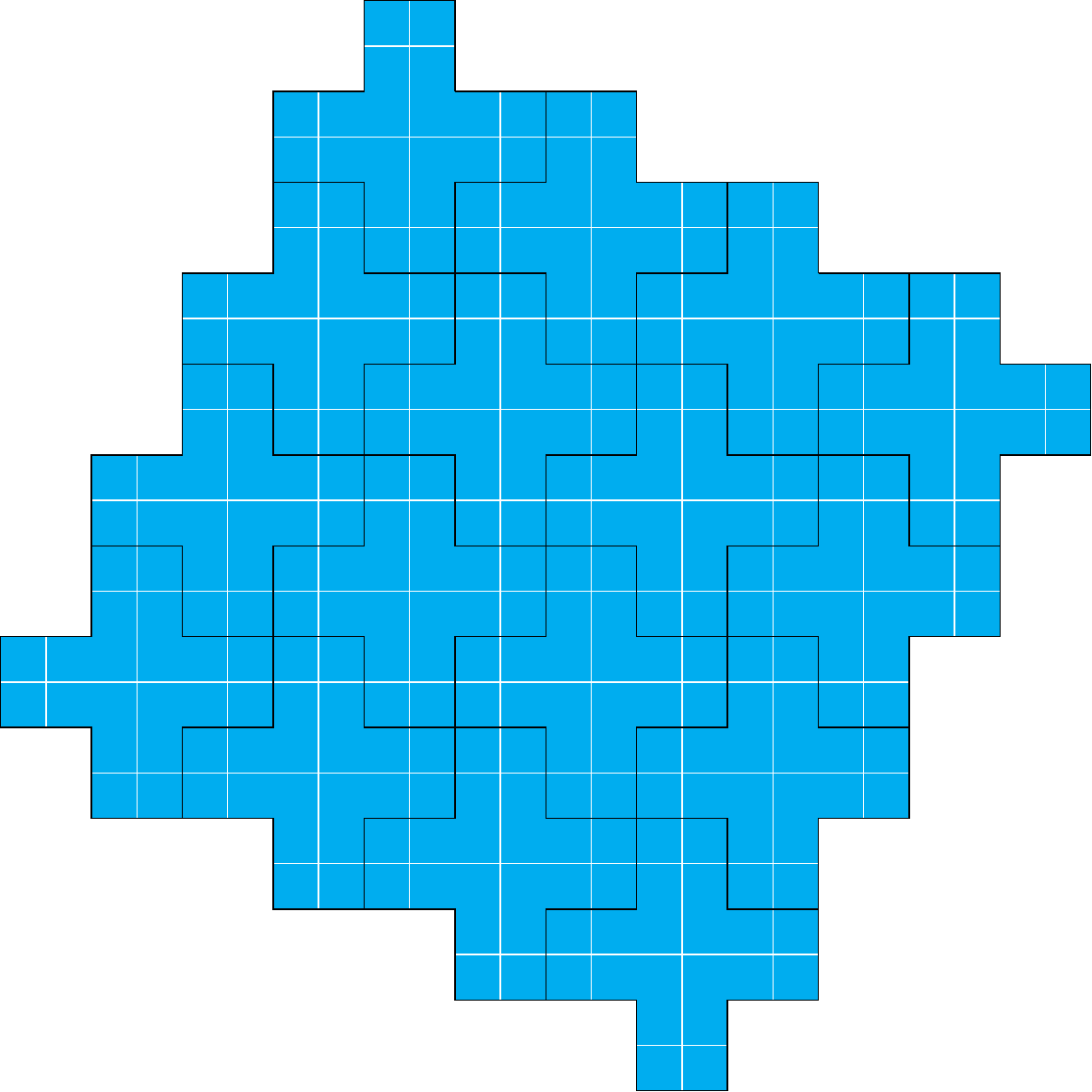}
  \hfil\includegraphics[width=6cm]{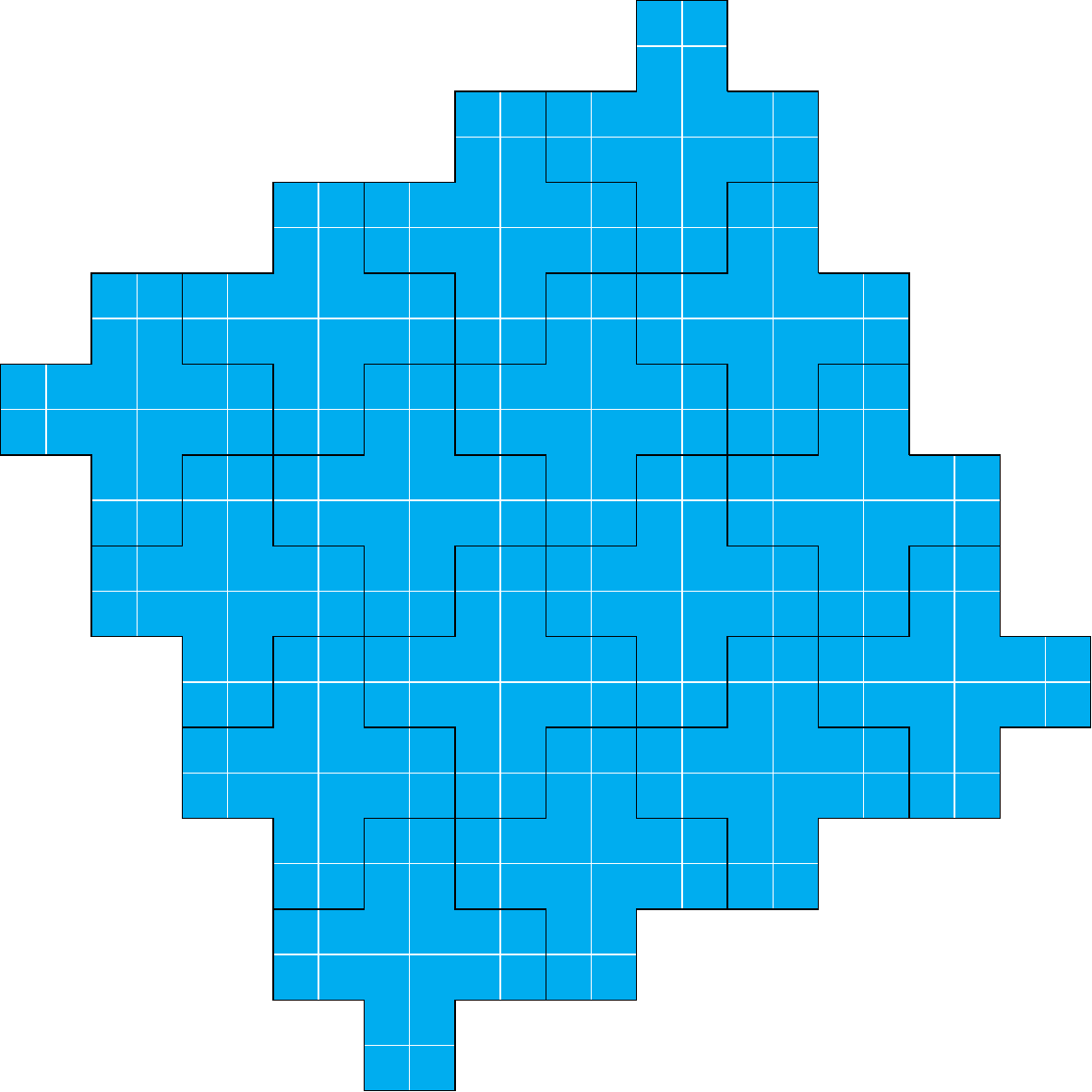}
  \caption{Perfect coverings of crosses. There are 10 inequivalent such coverings, obtained by translating each of the ones depicted here in 5 inequivalent ways. These two coverings are related to each other by a reflection.}
  \label{fig:cross_packing}
\end{figure}

\begin{figure}
  \hfil\includegraphics[width=2.5cm]{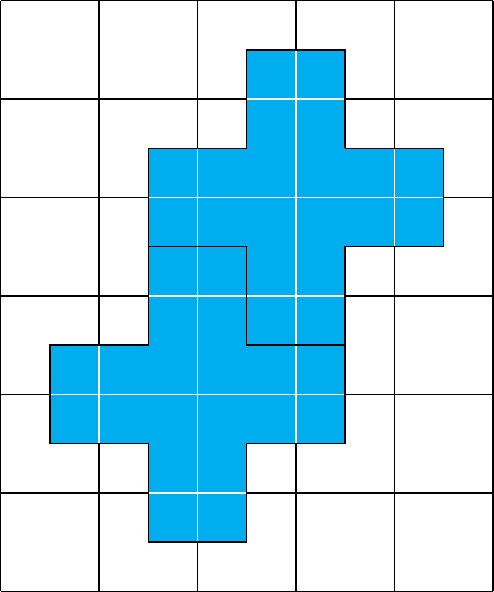} {\footnotesize\it a}.
  \hfil\includegraphics[width=3cm]{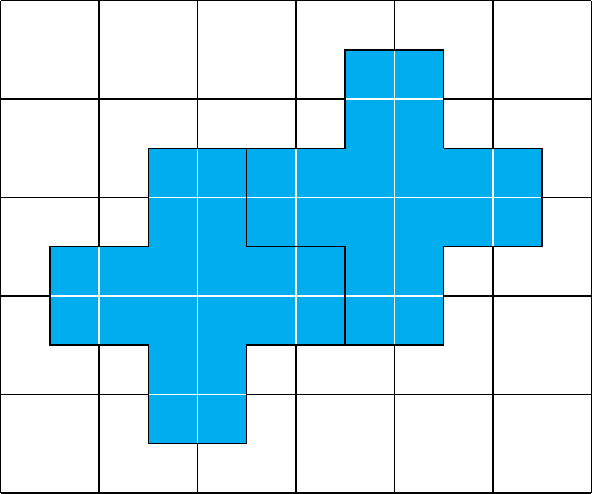} {\footnotesize\it b}.
  \hfil\includegraphics[width=3.5cm]{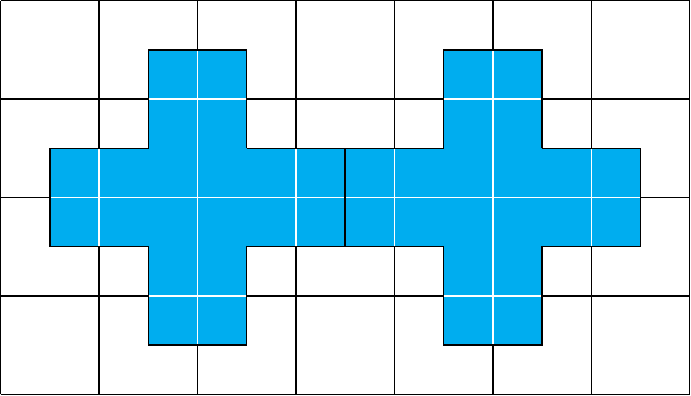} {\footnotesize\it c}.
  \caption{Pairs of crosses that are ({\it a}) left-packed, ({\it b}) right-packed and ({\it c}) stacked.}
  \label{fig:cross_pair_classify}
\end{figure}

\begin{figure}
  \hfil\includegraphics[width=3.5cm]{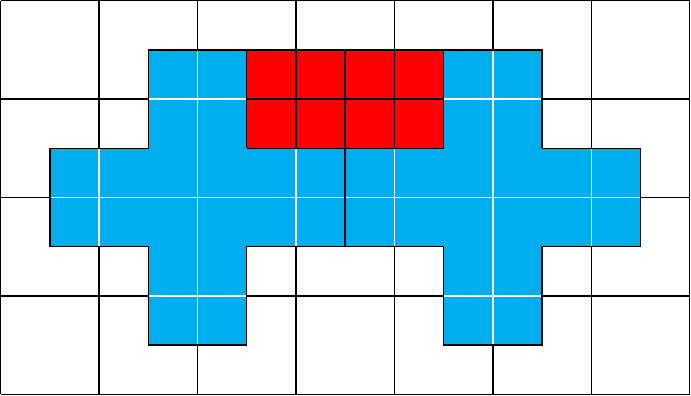} {\footnotesize\it a}.
  \hfil\includegraphics[width=3.5cm]{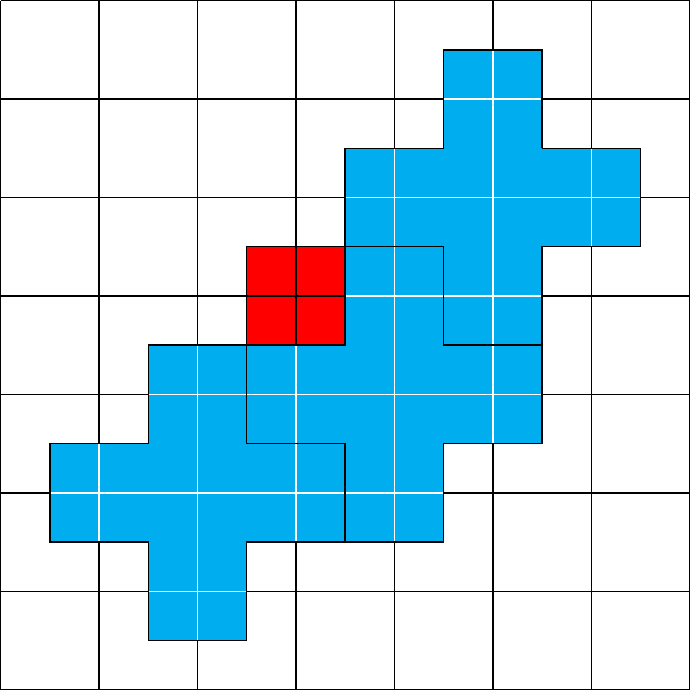} {\footnotesize\it b}.
  \hfil\includegraphics[width=4cm]{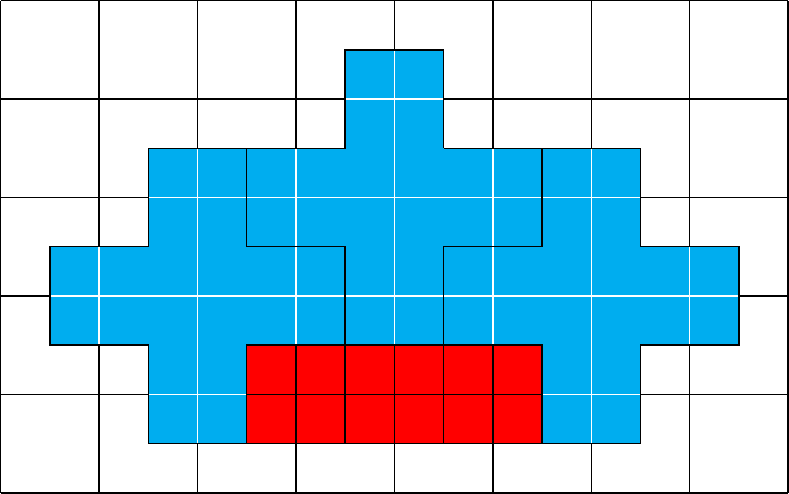} {\footnotesize\it c}.
  \caption{Connected configurations that cannot be completed to a perfect covering. The red (color online) regions cannot be entirely covered by crosses.}
  \label{fig:cross_sliding}
\end{figure}

\begin{figure}
  \hfil\includegraphics[width=4cm]{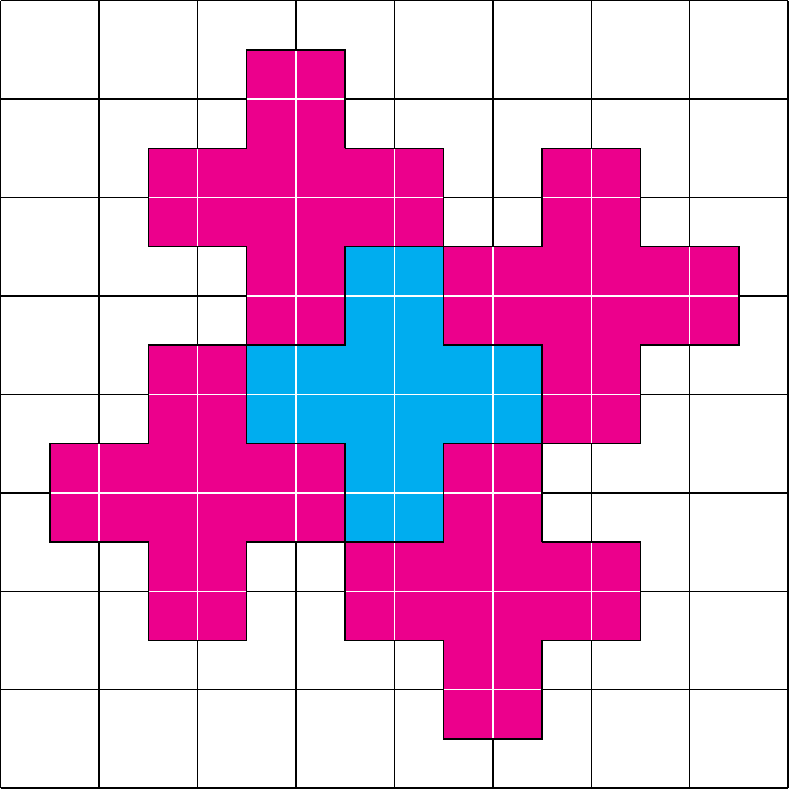} {\footnotesize\it a}.
  \hfil\includegraphics[width=4cm]{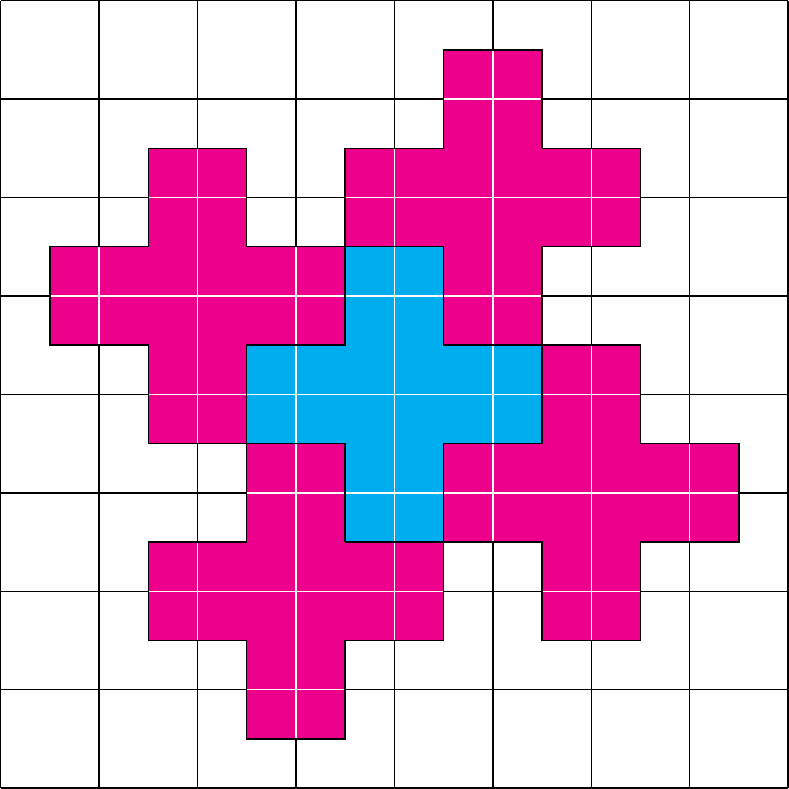} {\footnotesize\it b}.
  \caption{The two configurations in $\mathbb S(\{x\})\equiv\{X_a,X_b\}$. The cross at $x$ is drawn in cyan (color online), whereas the crosses in $X_i\setminus\{x\}$ are drawn in magenta (color online). For each $i\in\{a,b\}$, there exists a unique $\mu_i$ such that $X_i\subset\mathcal L_{\mu_i}$.}
  \label{fig:cross_unique1}
\end{figure}

\begin{figure}
  \hfil\includegraphics[width=4.5cm]{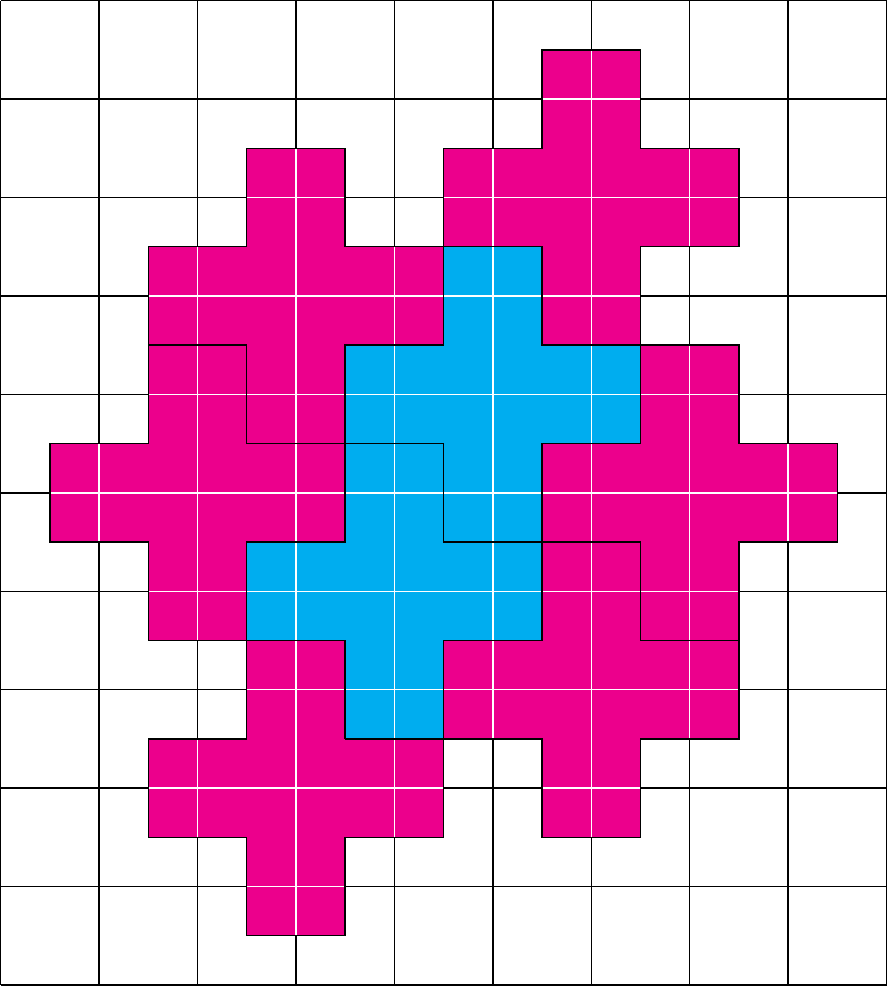}
  \caption{If $X$ is a pair of left-stacked crosses (in cyan, color online), then this is the unique configuration $X'\in\mathbb S(X)$. The crosses in $X'\setminus X$ are drawn in magenta (color online).}
  \label{fig:cross_unique2}
\end{figure}

\point Let us now consider the hard-cross model (see figure~\-\ref{fig:shapes}{\it b}), for which $\Lambda_\infty=\mathbb Z^2$, and
\begin{equation}
  \omega=\textstyle\left\{(n_x+x,n_y+y),\ (x,y)\in(-\frac12,\frac12)^2,\ (n_x,n_y)\in\{-1,0,1\}^2,\ |n_x|+|n_y|\leqslant 1\right\}.
  \label{shape_cross}
\end{equation}
There are $10$ perfect coverings (see figure~\-\ref{fig:cross_packing}):
\begin{equation}
  \mathcal L_1=\{(n_x+2n_y,2n_x-n_y),\ (n_x,n_y)\in\mathbb Z^2\}
  ,\quad
  \mathcal L_2=\{(-n_x+2n_y,2n_x+n_y),\ (n_x,n_y)\in\mathbb Z^2\}
  \label{coverings_cross}
\end{equation}
and, for $p\in\{2,3,4,5\}$,
\begin{equation}
  \mathcal L_{2p-1}=v_p+\mathcal L_1
  ,\quad
  \mathcal L_{2p}=v_p+\mathcal L_2
\end{equation}
with $v_2=(1,0)$, $v_3=(0,1)$, $v_4=(-1,0)$ and $v_5=(0,-1)$. The $\mathcal L_{2p-1}$ are related to $\mathcal L_1$ by translations, as are the $\mathcal L_{2p}$ related to $\mathcal L_2$, and $\mathcal L_2$ is mapped to $\mathcal L_1$ by the vertical reflection. Let us now check the non-sliding property. We first introduce the following definitions: two crosses at $x,x'$ whose supports are connected and disjoint are said to be (see figure~\-\ref{fig:cross_pair_classify})
\begin{itemize}
  \item {\it left-packed} if $x-x'\in\{(1,2),(-2,1),(-1,-2),(2,-1)\}\subset\mathcal L_1$
  \item {\it right-packed} if $x-x'\in\{(2,1),(-1,2),(-2,-1),(1,-2)\}\subset\mathcal L_2$
  \item {\it stacked} if $x-x'\in\{(3,0),(0,3),(-3,0),(0,-3)\}$.
\end{itemize}
Now, consider a connected configuration of crosses $X$.
\begin{itemize}
  \item If $|X|=1$, then $\mathbb S(X)$ (see definition~\-\ref{def:sliding}) consists of the two configurations in figure~\-\ref{fig:cross_unique1}, each of which is the subset of a unique sublattice $\mathcal L_\mu$.
  \item If $X$ contains at least one pair $x,x'\in X$ of stacked crosses, which, without loss of generality, we assume satisfies $x-x'=(-3,0)$, then one of the two sites $x+(1,1)$ or $x+(2,1)$ cannot be covered by any other cross (see figure~\-\ref{fig:cross_sliding}{\it a}), which implies that $\mathbb S(X)=\emptyset$.
  \item We now assume that every pair of crosses in $X$ is either left- or right-packed, and there exists at least one triplet $x,x',x''\in X$ whose supports are connected and disjoint, and is such that $x,x'$ is right-packed and $x,x''$ is left-packed. Without loss of generality, we assume that $x-x'=(2,1)$ and $x-x''=(-1,-2)$ (see figure~\-\ref{fig:cross_sliding}{\it b}) or $x-x''=(-2,1)$ (see figure~\-\ref{fig:cross_sliding}{\it c}). In the former case, the site $x+(-1,1)$ cannot be covered by any other crosses. In the latter case, one of the three sites $x+(-1,-2)$, $x+(0,-2)$ or $x+(1,-2)$ cannot be covered by any other cross. Thus, $\mathbb S(X)=\emptyset$.
  \item Finally, suppose that every pair of crosses is left-packed (the case in which they are all right-packed is treated identically). Let $Y$ be a pair of left-packed crosses, $\mathbb S(Y)$ consists of a single configuration, depicted in figure~\-\ref{fig:cross_unique2}, which is a subset of a unique sublattice $\mathcal L_\mu$. Since there is a unique way of isolating each left-packed pair in $X$, there is a single way of isolating $X$, that is, $\mathbb S(X)$ consists of a single configuration, which is the union over left-packed pairs $Y$ in $X$ of the unique configuration in $\mathbb S(Y)$, and is, therefore, a subset of a unique sublattice $\mathcal L_\mu$.
\end{itemize}
\bigskip

\begin{figure}
  \hfil\includegraphics[width=1.5cm]{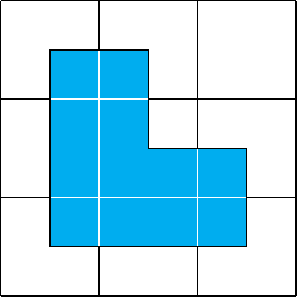}
  \hfil\includegraphics[width=1.5cm]{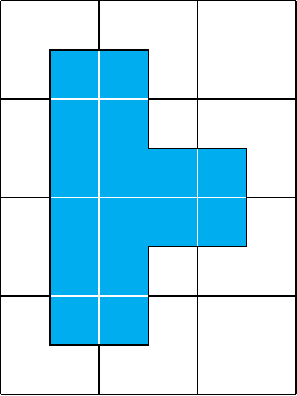}
  \hfil\includegraphics[width=1.5cm]{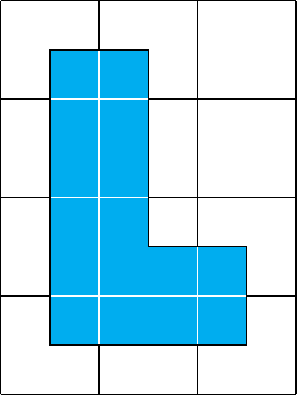}
  \hfil\includegraphics[width=1.5cm]{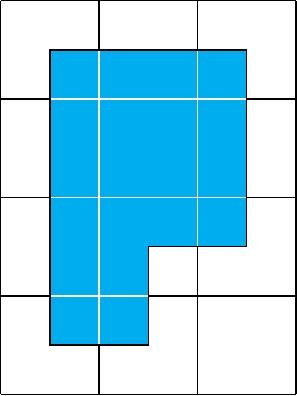}
  \caption{More examples of non-sliding hard-core lattice particle systems. These shapes are all {\it polyominoes}.}
  \label{fig:nonsliding_extra}
\end{figure}

\point By proceeding in a similar way, one proves that the models depicted in figure~\-\ref{fig:nonsliding_extra} are all non-sliding hard-core lattice particle systems. There are many more examples, among which the hard hexagon model (see figure~\-\ref{fig:shapes}{\it c}), and many more polyominoes than those depicted in figure~\-\ref{fig:nonsliding_extra}. In addition, for every hard polyomino model (a cross is a polyomino) that is non-sliding, the corresponding model with a finer lattice mesh is also non-sliding.
\bigskip

\section{High-fugacity expansion}

\indent In this section, we will prove the convergence of the high-fugacity expansion for non-sliding hard-core lattice particle systems. To that end, we will map the particle system to a model of {\it Gaunt-Fisher configurations} (GFc), and use a cluster expansion to compute the GFc partition function.
\bigskip

\subsection{The GFc model}
\indent We start by mapping the particle system to a model of Gaunt-Fisher configurations. This step is analogous to the contour mapping in the Peierls argument~\-\cite{Pe36}, which we will now briefly recall. Consider the two-dimensional ferromagnetic Ising model. Having fixed a boundary condition in which every spin on the boundary is up, one can represent any spin configuration as a collection of {\it contours}, which correspond to the interfaces of the regions of up and down spins. Since these boundaries are unlikely at low temperatures, the effective activity of a contour is low. We wish to adapt this construction to non-sliding hard-core lattice systems. Defining boundaries in this context is more delicate than in the Ising model, due to the necessity of constructing a model of contours that does not have any long range interactions. We will identify boundaries by focusing on empty space, and define GFcs as the connected components of the union of the empty space and the supports of the particles surrounding it. GFcs give us a formal way of defining the notion of a {\it defect}, which was left imprecise until now. The following definition follows somewhat naturally from the proof of lemma~\-\ref{lemma:GFc} below.
\bigskip

\theoname{Definition}{Gaunt-Fisher configurations}\label{def:GFc}
  Given $\nu\in\{1,\cdots,\tau\}$, a GFc is a quadruplet $\gamma\equiv(\Gamma_\gamma,X_\gamma,\nu,\underline\mu_\gamma)$ in which $\Gamma_\gamma$ is a {\it connected} and {\it bounded} subset of $\Lambda$, $X_\gamma\in\Omega(\Gamma_\gamma)$, and $\underline\mu_\gamma$ is a map $\mathcal H(\Gamma_\gamma)\to\{1,\cdots,\tau\}$, and satisfies the following condition. Let $\mathfrak X_\gamma$ denote the particle configuration obtained by covering the exterior and holes of $\Gamma_\gamma$ by particles:
  \begin{equation}
    \mathfrak X_\gamma:=
    \left(\mathcal L_\nu\cap\hat\Gamma_{\gamma,0}\right)\cup
    \left(\bigcup_{j=1}^{h_{\Gamma_\gamma}}\left(\mathcal L_{\underline\mu_\gamma(\hat\Gamma_{\gamma,j})}\cap\hat\Gamma_{\gamma,j}\right)\right)
    .
  \end{equation}
  A quadruplet $\gamma$ is a GFc if
  \begin{itemize}
    \item The particles in $X_\gamma$ are entirely contained inside $\Gamma_\gamma$ and those in $\mathfrak X_\gamma$ do not intersect $\Gamma_\gamma$: $\forall x\in X_\gamma$, $\sigma_x\subset\Gamma_\gamma$ and $\forall x'\in\mathfrak X_\gamma$, $\sigma_x\cap\Gamma_\gamma=\emptyset$.
    \item for every $x\in X_\gamma$, $\Delta(\sigma_x,\mathcal E_\Lambda(X_\gamma\cup\mathfrak X_\gamma))=1$ (recall that $\Delta$ is the graph distance on $\Lambda_\infty$, $\sigma_x$ is the support of the particle at $x$~\-(\ref{sigma}), and $\mathcal E_\Lambda(X_\gamma\cup\mathfrak X_\gamma)$ is the set of sites left uncovered by the configuration $X_\gamma\cup\mathfrak X_\gamma$~\-(\ref{mcE})),
    \item for every $x\in \mathfrak X_\gamma$, $\Delta(\sigma_x,\mathcal E_\Lambda(X_\gamma\cup\mathfrak X_\gamma))>1$.
  \end{itemize}
  We denote the set of GFcs by $\mathfrak C_\nu(\Lambda)$.
\endtheo
\bigskip

\theoname{Lemma}{GFc mapping}\label{lemma:GFc}
  The partition function~\-(\ref{Xi_nu}) can be rewritten as
  \begin{equation}
    \frac{\Xi^{(\nu)}_\Lambda(\underline z)}{\mathbf z_\nu(\Lambda)}=
    \sum_{\underline\gamma\subset\mathfrak C_\nu(\Lambda)}
    \left(\prod_{\gamma\neq\gamma'\in\underline\gamma}\Phi(\gamma,\gamma')\right)
    \prod_{\gamma\in\underline\gamma}\zeta_\nu^{(\underline z)}(\gamma)
    \label{XiGFc}
  \end{equation}
  where $\mathfrak C_\nu(\Lambda)$ is the set of GFcs, defined in definition~\-\ref{def:GFc} below, $\Phi(\gamma,\gamma')\in\{0,1\}$ is equal to 1 if and only if $\Gamma_\gamma$ and $\Gamma_{\gamma'}$ are disconnected,
  \begin{equation}
    \mathbf z_\nu(\Lambda):=\prod_{x\in\Lambda\cap\mathcal L_\nu}z(x)
    \label{bfz}
  \end{equation}
  and
  \begin{equation}
    \zeta_\nu^{(\underline z)}(\gamma):=
    \frac
      {\prod_{x\in X_\gamma}z(x)}
      {\mathbf z_\nu(\Gamma_\gamma)}
    \prod_{j=1}^{h_{\Gamma_\gamma}}
    \frac
      {\Xi_{\hat\Gamma_{\gamma,j}}^{(\underline\mu_\gamma(\hat\Gamma_{\gamma,j}))}(\underline z)}
      {\Xi_{\hat\Gamma_{\gamma,j}}^{(\nu)}(\underline z)}
    \label{zeta}
  \end{equation}
  in which we used the following definition. Given a connected subset $\Gamma\subset\Lambda$, we denote the {\it exterior} of $\Gamma$ by $\hat\Gamma_0$, and its {\it holes} by $\mathcal H(\Gamma)\equiv\{\hat\Gamma_1,\cdots,\hat\Gamma_{h_{\Gamma}}\}$ with $h_{\Gamma}\geqslant 0$. Formally, $\hat\Gamma_0,\cdots,\hat\Gamma_{h_\Gamma}$ are the connected components of $\Lambda_\infty\setminus\Gamma$, and $\hat\Gamma_0$ is the only unbounded one.
\endtheo
\bigskip

\indent\underline{Proof}:
  We will first map particle configurations to a set of GFc, then extract the most external ones, and conclude the proof by induction.
  \bigskip

\begin{figure}
    \hfil\includegraphics[width=14cm]{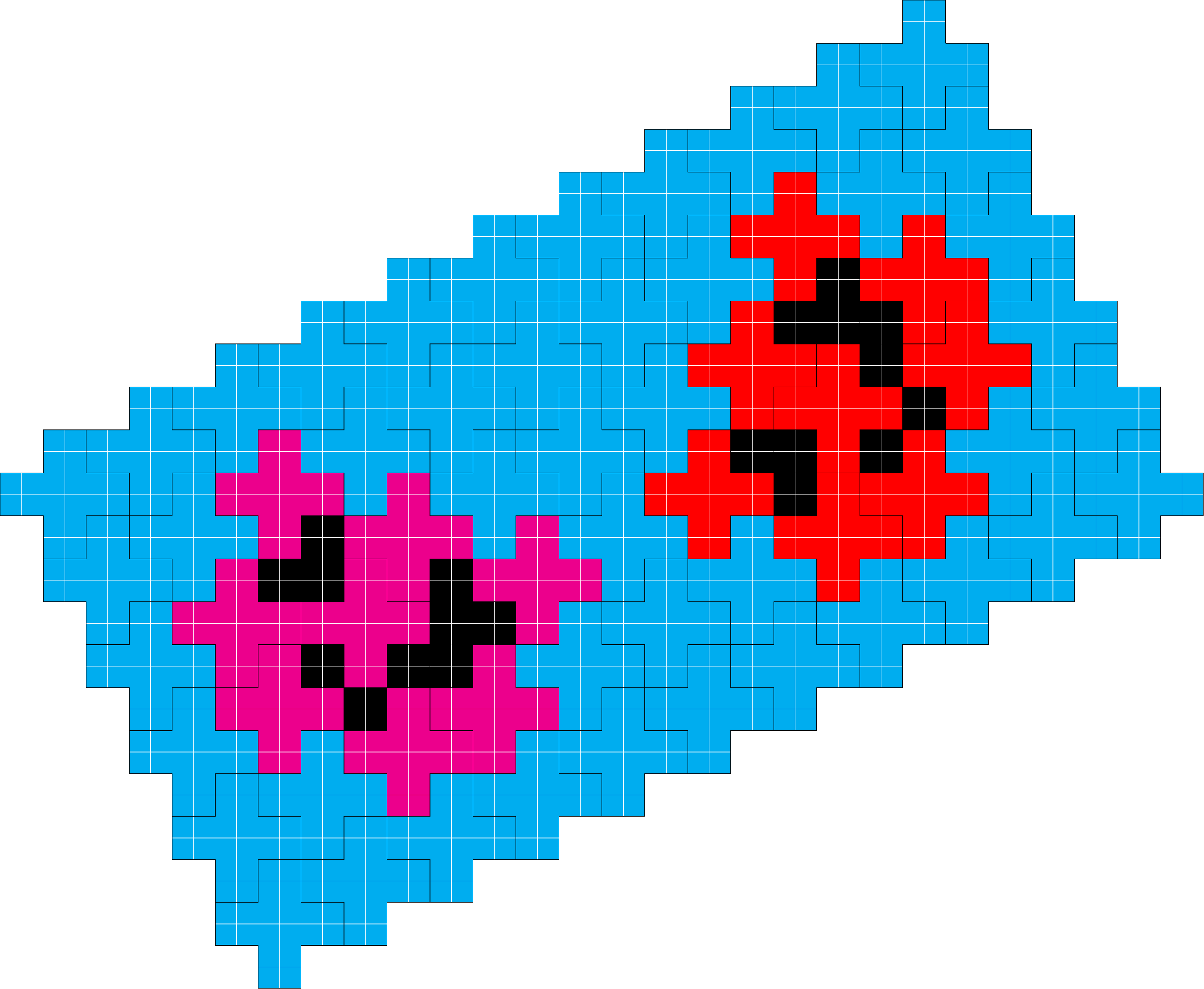}
    \caption{An example cross configuration, and its associated GFc supports. There are two disconnected GFcs: the first consists of the red (color online) crosses and the neighboring black empty sites, and the second consists of the magenta (color online) crosses and the neighboring black empty sites.}
    \label{fig:contour_example}
  \end{figure}

  \point{\bf GFcs.} To a configuration $X\in\Omega_\nu(\Lambda)$, we associate a set of {\it external GFcs}. See figure~\-\ref{fig:contour_example} for an example.
  \bigskip

  \indent Given $x\in\Lambda$, let $\partial_X(x)$ denote the set of sites covered by particles neighboring $x$ which do not themselves cover $x$:
  \begin{equation}
    \partial_X(x):=\bigcup_{\displaystyle\mathop{\scriptstyle y\in X}_{\Delta(\sigma_y,x)=1}}\sigma_y.
  \end{equation}
  Consider the union of the set of empty sites and the particles neighboring it:
  \begin{equation}
    \mathbb U_\Lambda(X):=\mathcal E_\Lambda(X)\cup\left(\bigcup_{x\in\mathcal E_\Lambda(X)}\partial_X(x)\right)
  .
  \end{equation}
  We denote the connected components of $\mathbb U_\Lambda(X)$ by $\Gamma_1,\cdots,\Gamma_n$. These will be the supports of the GFcs associated to the configuration.
  \bigskip

\begin{figure}
    \hfil\includegraphics[width=14cm]{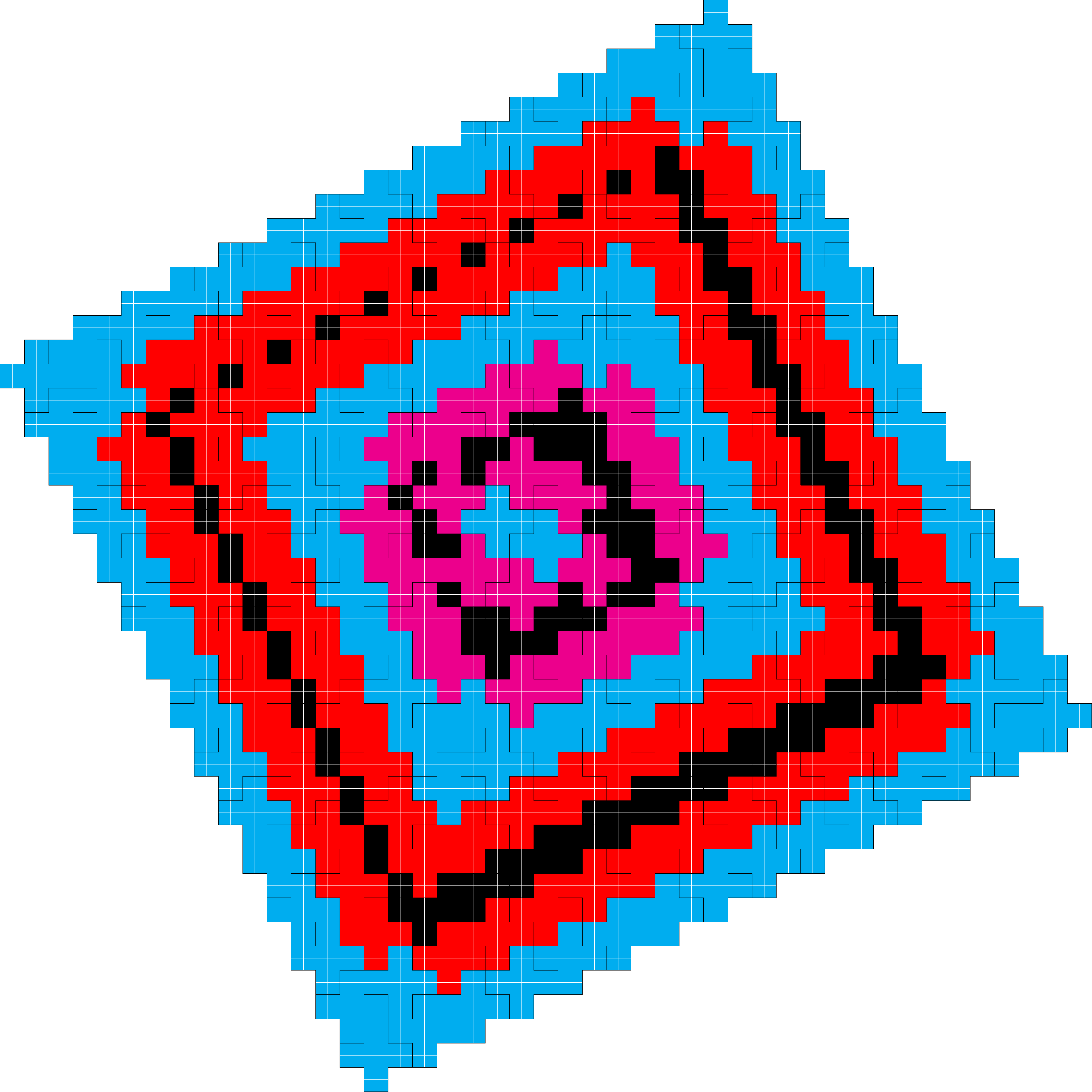}
    \caption{A configuration in which the GFc supports are nested. The $\kappa_i$ are the connected components of cyan (color online) crosses. Each is a subset of a unique perfect covering.}
    \label{fig:contour_nested}
  \end{figure}

  \indent We then denote the connected components of $\Lambda_\infty\setminus(\Gamma_1\cup\cdots\cup\Gamma_n)$ by $\{\kappa_1,\cdots,\kappa_m\}$. By construction, each $\kappa_i$ is covered by particles. We denote the particle configuration restricted to $\kappa_i$ by $X_i:=X\cap\kappa_i$. In addition, we define $\bar X_i$ as the union of $X_i$ and the particles that surround $\kappa_i$:
  \begin{equation}
    \bar X_i:=X_i\cup\{x\in X,\ \exists x'\in X_i,\ \Delta(\sigma_x,\sigma_{x'})=1\}\in\mathbb S(X_i)
  \end{equation}
  (we recall that $\mathbb S$ was defined in definition~\-\ref{def:sliding}). By the non-sliding condition, there exists a {\it unique} $\mu_i\in\{1,\cdots,\tau\}$ such that $\bar X_i\subset\mathcal L_{\mu_i}$. See figure~\-\ref{fig:contour_nested} for an example.
  \bigskip

  \indent By construction, for every $i\in\{1,\cdots,n\}$, each hole of $\Gamma_i$ (we recall that the holes of $\Gamma_i$ are denoted by $\hat\Gamma_{i,j}$) contains at least one of the $\kappa_k$. In fact, for every $i\in\{1,\cdots,n\}$ and $j\in\{0,\cdots,h_{\Gamma_i}\}$ there exists a unique index $k(\hat\Gamma_{i,j})\in\{1,\cdots,m\}$ such that $\kappa_{k(\hat\Gamma_{i,j})}$ is contained inside $\hat\Gamma_{i,j}$ and is in contact with $\Gamma_i$:
  \begin{equation}
    \kappa_{k(\hat\Gamma_{i,j})}\subset\hat\Gamma_{i,j}
    ,\quad
    \Delta(\kappa_{k(\hat\Gamma_{i,j})},\Gamma_i)=1
  \end{equation}
  (see figure~\-\ref{fig:contour_nested}). We then define the set of GFcs associated to $X$ as the set of quadruplets
  \begin{equation}
    \underline\gamma(X)=\left\{\left(\Gamma_i,X\cap\Gamma_i,\ \mu_{k(\hat\Gamma_{i,0})},\ \underline\mu_i\right),\quad i\in\{1,\cdots,n\}\right\}
    \label{GFcs_X}
  \end{equation}
  where $X\cap\Gamma_i$ is the restriction of the particle configuration to $\Gamma_i$, and $\underline\mu_i$ is the map from $\mathcal H(\hat\Gamma_i)$ to $\{1,\cdots,\tau\}$ defined by
  \begin{equation}
    \underline\mu_i(\hat\Gamma_{i,j})=\mu_{k(\hat\Gamma_{i,j})}.
  \end{equation}
  The set of quadruplets thus constructed is a set of GFcs, in the sense of definition~\-\ref{def:GFc}, that is, $\underline\gamma(X)\subset\mathfrak C_\nu(\Lambda)$.
  \bigskip

  \point{\bf External GFc model.} We have thus mapped $X$ to a model of GFcs. Note that the indices $\mu_\cdot$ must match up, that is, if a GFc is the first nested GFc in the hole of another, its external $\mu$ must be equal to the $\mu$ of the hole it is in. This is a long range interaction between GFcs, which makes the GFc model difficult to study. Instead, we will map the system to a model of {\it external} GFcs, that do not have long range interactions. We introduce the following definitions: two GFcs $\gamma,\gamma'\in\mathfrak C_\nu(\Lambda)$ are said to be
  \begin{itemize}
    \item {\it compatible} if their supports are disconnected, that is, $\Delta(\Gamma_{\gamma},\Gamma_{\gamma'})>1$,
    \item {\it external} if their supports are in each other's exteriors, that is, $\Gamma_{\gamma}\subset\hat\Gamma_{\gamma',0}$ and $\Gamma_{\gamma'}\subset\hat\Gamma_{\gamma,0}$.
  \end{itemize}
  The GFcs in $\underline\gamma(X)$ (see~\-(\ref{GFcs_X})) are compatible, but not necessarily external to each other. Roughly, the idea is to keep the GFcs that are external to each other, since those do not have long-range interactions (they all share the same external $\mu$, which is fixed to $\nu$ once and for all). At that point, the particle configuration in the exterior of all GFcs is fixed, and we are left with summing over configurations in the holes. The sum over configurations in each hole is of the same form as~\-(\ref{Xi_nu}), with $\Lambda$ replaced by the hole, and the boundary condition by the appropriate $\underline\mu$. Following this, we rewrite~\-(\ref{Xi_nu}) as
  \begin{equation}
    \frac{\Xi^{(\nu)}_\Lambda(\underline z)}{\mathbf z_\nu(\Lambda)}=
    \sum_{\underline\gamma\subset\mathfrak C_\nu(\Lambda)}
    \left(\prod_{\gamma\neq\gamma'\in\underline\gamma}\Phi_{\mathrm{ext}}(\gamma,\gamma')\right)
    \prod_{\gamma\in\underline\gamma}
    \left(
      \frac{\prod_{x\in X_\gamma}z(x)}{\mathbf z_\nu(\Gamma_\gamma)}
      \prod_{j=1}^{h_{\Gamma_\gamma}}
      \frac{\Xi_{\hat\Gamma_{\gamma,j}}^{(\underline\mu_\gamma(\hat\Gamma_{\gamma,j}))}(\underline z)}{\mathbf z_\nu(\hat\Gamma_{\gamma,j})}
    \right)
    \label{Xiexternal}
  \end{equation}
  in which $\Phi_{\mathrm{ext}}(\gamma,\gamma')\in\{0,1\}$ is equal to 1 if and only if $\gamma$ and $\gamma'$ are {\it compatible} and {\it external}. Note that $\hat\Gamma_{\gamma,j}$ is obviously bounded, connected and $\Lambda_\infty\setminus\hat\Gamma_{\gamma,j}$ is connected. It is also tiled, since, as is readily checked,
  \begin{equation}
    \hat\Gamma_{\gamma,j}
    =
    \bigcup_{x\in\mathcal L_{\underline\mu_i(\hat\Gamma_{\gamma,j})}\cap\hat\Gamma_{\gamma,j}}\sigma_x
    .
  \end{equation}
  
  We have, thus, rewritten the model as a system of external GFcs.
  \bigskip

  \point{\bf GFc model.} The last factor in~\-(\ref{Xiexternal}) is similar to the left side of~\-(\ref{Xiexternal}), except for the fact that the boundary condition is $\underline\mu_\gamma(\hat\Gamma_{\gamma,j})$ instead of $\nu$. (The denominator $\mathbf z_\nu$ also has a different index from the numerator, although this is not a problem since $\mathbf z_\nu$ and $\mathbf z_{\underline\mu_\gamma}$ are rather explicit.) In order to obtain a model of GFcs (which are not necessarily external to each other), we could iterate~\-(\ref{Xiexternal}), but, as was discussed earlier, this would induce long-range correlations. Instead, we introduce a trivial identity into~\-(\ref{Xiexternal}):
  \begin{equation}
    \frac{\Xi^{(\nu)}_\Lambda(\underline z)}{\mathbf z_\nu(\Lambda)}=
    \sum_{\underline\gamma\subset\mathfrak C_\nu(\Lambda)}
    \left(\prod_{\gamma\neq\gamma'\in\underline\gamma}\Phi_{\mathrm{ext}}(\gamma,\gamma')\right)
    \prod_{\gamma\in\underline\gamma}
    \left(
      \zeta_\nu^{(\underline z)}(\gamma)
      \prod_{j=1}^{h_{\Gamma_\gamma}}
      \frac{\Xi_{\hat\Gamma_{\gamma,j}}^{(\nu)}(\underline z)}{\mathbf z_\nu(\hat\Gamma_{\gamma,j})}
    \right)
    \label{Xiexternal2}
  \end{equation}
  in which $\zeta_\nu^{(\underline z)}(\gamma)$ is defined in~\-(\ref{zeta}). We then rewrite $\Xi_{\hat\Gamma_{\gamma,j}}^{(\nu)}(\underline z)$ using~\-(\ref{Xiexternal2}), iterate, and, noting that, if $\hat\Gamma_{\gamma,j}$ does not contain GFcs, then $\Xi_{\hat\Gamma_{\gamma,j}}^{(\nu)}(\underline z)=\mathbf z_\nu(\hat\Gamma_{\gamma,j})$, we find~\-(\ref{XiGFc}).
\qed

\subsection{Cluster expansion of the GFc model}
\indent As was discussed in section~\-\ref{sec:low_fugacity}, the pressure of a system of hard particles at low fugacity can be expressed as a convergent power series. The GFc model in~\-(\ref{XiGFc}) is a system of hard GFcs (the factor $\Phi(\gamma,\gamma')$ is a hard-core interaction), and, as we will see below, the GFcs have a small activity. Similarly to the low-fugacity expansion, the logarithm of the left side of~\-(\ref{XiGFc}) can be expressed as a convergent power series. In this context, in which the hard GFcs have more structure than hard particles, the expansion is usually called a {\it cluster expansion}. The cluster expansion has been studied extensively (to cite but a few~\-\cite{Ru99,GBG04,KP86,BZ00}), and we will use a theorem by Bovier and Zahradnik~\-\cite{BZ00}, which is summarized in the following lemma.
\bigskip

\theoname{Lemma}{convergence of the cluster expansion {\rm \cite{BZ00}}}\label{lemma:cluster_expansion}
  If there exist two functions $a,d$ that map $\mathfrak C_\nu(\Lambda)$ to $[0,\infty)$ and a number $\delta\geqslant0$, such that $\forall\gamma\in\mathfrak C_\nu(\Lambda)$,
  \begin{equation}
    |\zeta_\nu^{(\underline z)}(\gamma)|e^{a(\gamma)+d(\gamma)}\leqslant\delta<1
    ,\quad
    \sum_{\displaystyle\mathop{\scriptstyle\gamma'\in\mathfrak C_\nu(\Lambda)}_{\gamma'\not\sim\gamma}}|\zeta_\nu^{(\underline z)}(\gamma')|e^{a(\gamma')+d(\gamma')}\leqslant
    \frac{\delta}{|\log(1-\delta)|}
    a(\gamma)
    \label{cvcd}
  \end{equation}
  in which $\gamma'\not\sim\gamma$ means that $\gamma'$ and $\gamma$ are {\it not} compatible (that is, the union of their supports is connected), then
  \begin{equation}
    \frac{\Xi_\Lambda^{(\nu)}(\Lambda)}{\mathbf z_\nu(\Lambda)}
    =\exp\left(
      \sum_{\underline\gamma\sqsubset\mathfrak C_\nu(\Lambda)}
      \Phi^T(\underline\gamma)
      \prod_{\gamma\in\underline\gamma}\zeta_\nu^{(\underline z)}(\gamma)
    \right)
    \label{ce}
  \end{equation}
  $\underline\gamma\sqsubset\mathfrak C_\nu(\Lambda)$ means that $\underline\gamma$ is a multiset (a multiset is similar to a set except for the fact that an element may appear several times in a multiset, in other words, a multiset is an unordered tuple) with elements in $\mathfrak C_\nu(\Lambda)$, and $\Phi^T$ is the {\it Ursell function}, defined as
  \begin{equation}
    \Phi^T(\gamma_1,\cdots,\gamma_n):=
    \frac1{N_{\underline\gamma}!}
    \sum_{\mathfrak g\in\mathcal G^T(n)}\prod_{\{j,j'\}\in\mathcal E(\mathfrak g)}(\Phi(\gamma_j,\gamma_{j'})-1)
    \label{PhiT}
  \end{equation}
  where $\Phi(\gamma_j,\gamma_{j'})\in\{0,1\}$ is equal to 1 if and only if $\Gamma_{\gamma_j}\cup\Gamma_{\gamma_{j'}}$ is {\it disconnected}, $\mathcal G^T(n)$ is the set of connected graphs on $n$ vertices and $\mathcal E(\mathfrak g)$ is the set of edges of $\mathfrak g$, and, if $n_{\gamma_i}$ is the multiplicity of $\gamma_i$ in $(\gamma_1,\cdots,\gamma_n)$, then $N_{\underline\gamma}!\equiv\prod_{j=1}^n(n_{\gamma_j}!)^{\frac1{n_{\gamma_j}}}$. In addition, for every $\gamma\in\mathfrak C_\nu(\Lambda)$,
  \nopagebreakaftereq
  \begin{equation}
    \sum_{\underline\gamma'\sqsubset\mathfrak C_\nu(\Lambda)}
    \left|
      \Phi^T(\{\gamma\}\sqcup\underline\gamma')
      \prod_{\gamma'\in\underline\gamma'}
      \left(\zeta_\nu^{(\underline z)}(\gamma')e^{d(\gamma')}\right)
    \right|
    \leqslant
    e^{a(\gamma)}
    \label{ce_remainder}
  \end{equation}
  where $\sqcup$ denotes the union operation in the sense of multisets.
\endtheo
\restorepagebreakaftereq
\bigskip

\indent We will now show that~\-(\ref{cvcd}) holds for an appropriate choice of $a$, $d$ and $\delta$.
\bigskip

\theoname{Lemma}{bound on the activity}\label{lemma:bound_zeta}
  Let
  \begin{equation}
    \mathcal N:=\sup_{x\in\Lambda_\infty,X\in\Omega(\Lambda_\infty)}|\partial_X(x)|.
    \label{mcN}
  \end{equation}
  If $z(x)\equiv z$ for every $x\in\Lambda_\infty$ except for a finite number $\mathfrak n$ of sites $(\mathfrak x_1,\cdots,\mathfrak x_{\mathfrak n})$, and if there exist $z_0,\cst c{cst:z}>0$ such that $|z|>z_0$ and
  \begin{equation}
    e^{-\frac{\cst c{cst:z}}{\mathfrak n}}|z|\leqslant |z(\mathfrak x_i)|\leqslant e^{\frac{\cst c{cst:z}}{\mathfrak n}}|z|
  \end{equation}
  then, for every $\theta,\xi\in(0,1)$ such that $\theta+\xi<1$, (\ref{cvcd}) is satisfied with
  \begin{equation}
    a(\gamma):=-\theta|\Gamma_\gamma|\log\alpha>0
    ,\quad
    d(\gamma):=-\xi|\Gamma_\gamma|\log\alpha>0
    \label{a}
  \end{equation}
  and
  \begin{equation}
    \delta=\varsigma\alpha^{1-(\theta+\xi)}
    ,\quad
    \varsigma=\mathrm{max}\left(e^{2\cst c{cst:z}},\ 1+2\mathfrak n(e^{2\frac{\cst c{cst:z}}{\mathfrak n}}+1)\right).
    \label{delta_sigma}
  \end{equation}
  in which
  \begin{equation}
    \alpha:=\varsigma e^{\chi}|z|^{-\rho_m(1+\mathcal N)^{-1}}\ll1
    \label{alpha}
  \end{equation}
  in which $\chi$ is the coordination number of $\Lambda_\infty$, that is, the maximal number of neighbors each vertex in $\Lambda_\infty$ has.
  \bigskip

  In addition, there exists $\cst C{cst:deriv_Xi}\in(0,\xi)$ such that, for every $i\in\{1,\cdots,\mathfrak n\}$, and every $\mu\in\{1,\cdots,\tau\}$
  \begin{equation}
    \left|\frac\partial{\partial\log z(\mathfrak x_i)}\log\left(
      \frac{\Xi_\Lambda^{(\mu)}(\underline z)}{\mathbf z_\mu(\Lambda)}
    \right)\right|
    \leqslant
    \alpha^{\cst C{cst:deriv_Xi}}\mathds 1(\mathfrak x_i\in\Lambda)
    \label{bound_deriv}
  \end{equation}
  in which $\mathds 1(E)\in\{0,1\}$ is equal to 1 if and only if $E$ is true.
\endtheo
\bigskip

{\bf Remark}: The value of $z_0$ depends on the model. It is worked out rather explicitly in the proof, and appears as a smallness condition on $\alpha$, which is made explicit in~\-(\ref{assum_alpha1}), (\ref{assum_alpha2}), (\ref{assum_alpha3}), (\ref{assum_alpha4}), (\ref{assum_alpha5}) and~\-(\ref{assum_alpha6}). In these equations, we use the notation $\alpha\ll(\cdots)$ to mean ``there exists a small constant $c>0$ such that if $\alpha<c(\cdots)$''.
\bigskip

\begin{figure}
    \hfil\includegraphics[width=7.5cm]{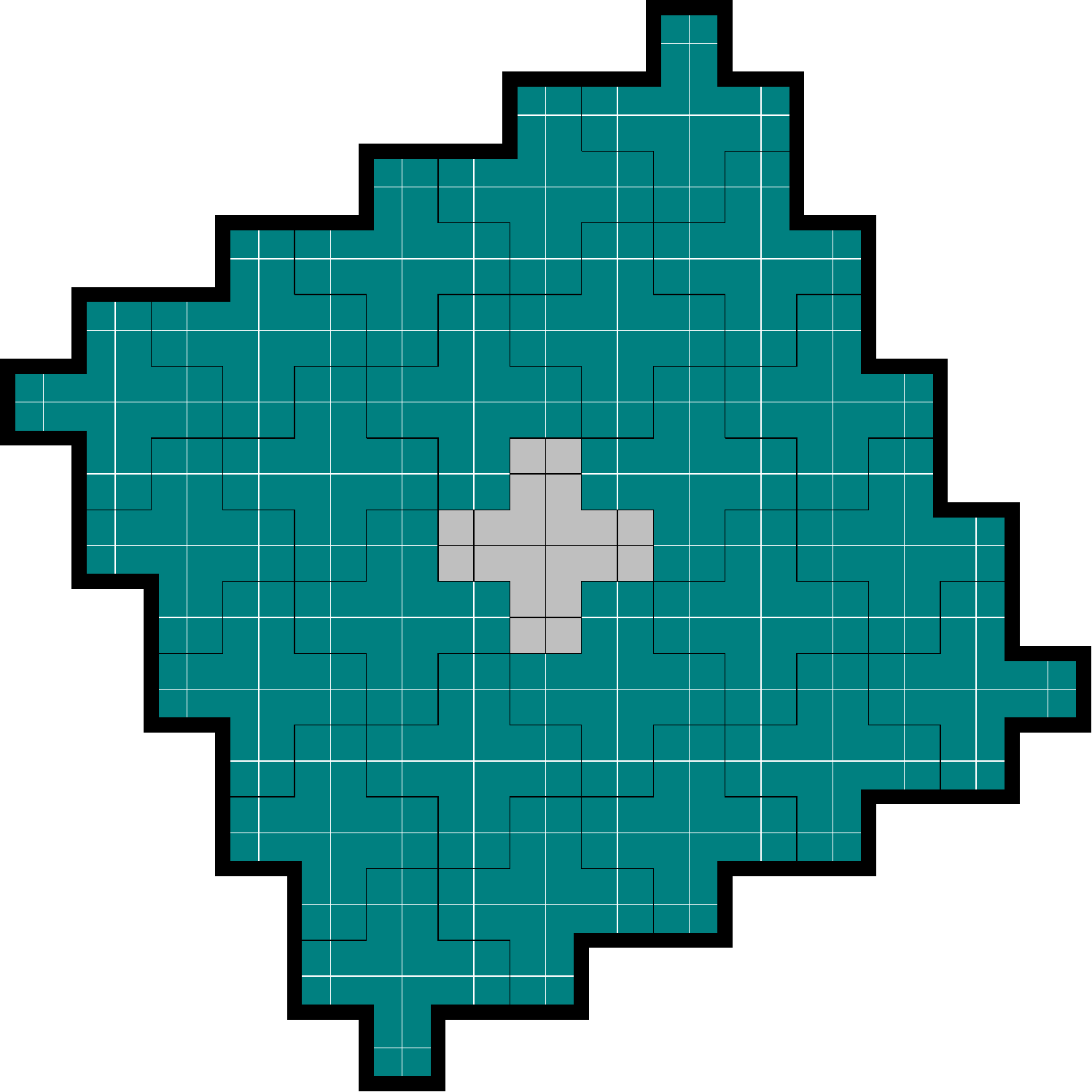}{\footnotesize\it a.}
    \hfil\includegraphics[width=7.5cm]{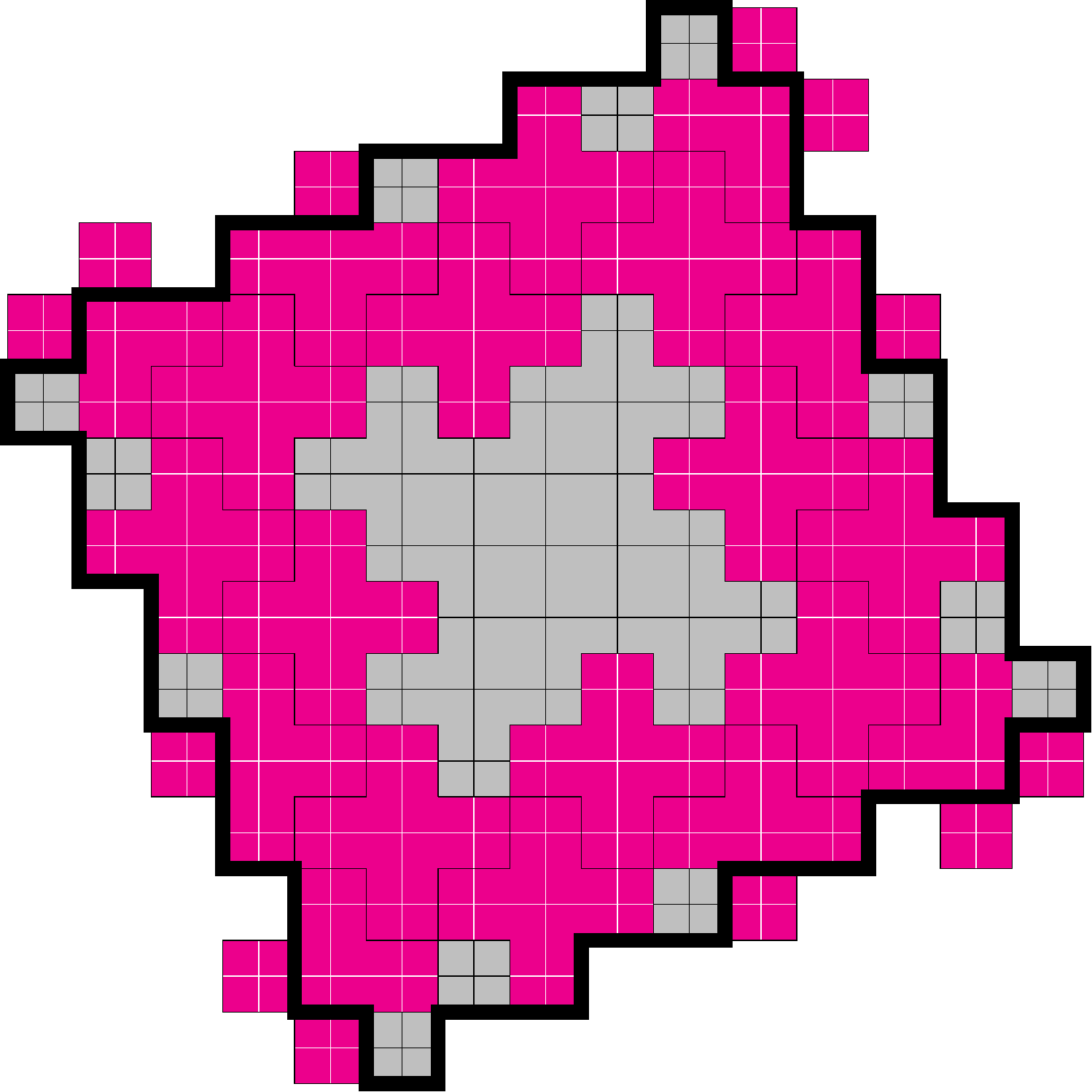}{\footnotesize\it b.}
    \caption{%
      Two different boundary conditions for the hard-cross model. The set $\Lambda$ is outlined by the thick black line. The crosses that are drawn are those mandated by the boundary condition (the boundary condition stipulates that every cross that is in contact with the boundary must be of a specified phase and cannot be in contact with empty sites), and the remaining available space in $\Lambda$ is colored gray. In figure~\-{\it a}, $\Lambda$ can be tiled by the covering corresponding to the boundary condition, whereas it cannot in figure~\-{\it b}. The partition function in the case of figure~\-{\it a} is
      $$z^{25}(1+y)$$
      whereas that in figure~\-{\it b} is
      $$z^{25}(1+5y+14y^2+18y^3+9y^4+y^5).$$
    }
    \label{fig:assymmetry}
  \end{figure}

\indent\underline{Proof}:
  We will prove this lemma along with the following inequality: for every $\mu\in\{1,\cdots,\tau\}$
  \begin{equation}
    \left|\frac{\Xi_{\Lambda}^{(\mu)}(z)}{\Xi_{\Lambda}^{(\nu)}(z)}\right|
    \leqslant
    \varsigma e^{|\partial\Lambda|}
    \label{assumK}
  \end{equation}
  in which $\partial\Lambda$ is the set of sites in $\Lambda$ that neighbor $\Lambda_\infty\setminus\Lambda$. We proceed by induction on the volume $|\Lambda|$ of $\Lambda$. (Note that, for certain models, this ratio is identically equal to 1. This is the case when the different perfect coverings are related to each other by a translation, as in the hard diamond model. However, for the hard-cross model, in which certain perfect coverings are related by a reflection, the ratio may differ from 1, see figure~\-\ref{fig:assymmetry}.)
  \bigskip

  \point First of all, if $\Lambda$ is so small that it cannot contain a GFc, that is, $\mathfrak C_\mu(\Lambda)=\emptyset$ for every $\mu\in\{1,\cdots,\tau\}$, then~\-(\ref{cvcd}) is trivially true, and
  \begin{equation}
    \Xi_{\Lambda}^{(\mu)}(\underline z)=\mathbf z_\mu(\Lambda)=\prod_{x\in\Lambda\cap\mathcal L_\mu}z(x)
    .
  \end{equation}
  Therefore, (\ref{bound_deriv}) holds. We now turn to~\-(\ref{assumK}). The $x$ dependence of $z(x)$ can be neglected, since there can be at most $\mathfrak n$ factors that differ from $z$, and they do so by a bounded amount:
  \begin{equation}
    e^{-\cst c{cst:z}}|z|^{|\Lambda\cap\mathcal L_\mu|}
    \leqslant
    |\Xi_{\Lambda}^{(\mu)}(\underline z)|
    \leqslant
    e^{\cst c{cst:z}}|z|^{|\Lambda\cap\mathcal L_\mu|}.
  \end{equation}
  In addition, as we will show below, $|\Lambda\cap\mathcal L_\mu|$ is independent of $\mu$, which implies that
  \begin{equation}
    \left|\frac{\Xi_{\Lambda}^{(\mu)}(\underline z)}{\Xi_{\Lambda}^{(\nu)}(\underline z)}\right|
    \leqslant e^{2\cst c{cst:z}}\leqslant \varsigma e^{|\partial\Lambda|}
    \label{assumK_0}
  \end{equation}
  since, by~\-(\ref{delta_sigma}),
  \begin{equation}
    \varsigma\geqslant e^{2\cst c{cst:z}}
    .
    \label{bound_varsigma}
  \end{equation}
  So, to conclude this argument, it suffices to prove that $|\Lambda\cap\mathcal L_\mu|$ is independent of $\mu$. This follows from the fact that $\Lambda$ is {\it tiled} (see~\-(\ref{tiled})). In fact, we will show that for every $x\in\Lambda_\infty$, $|\mathcal L_\mu\cap\sigma_x|=1$ for any $\mu$, which, by~\-(\ref{tiled}) implies that $|\Lambda\cap\mathcal L_\mu|=\rho_m|\Lambda|$. We proceed in two steps, by first showing that $|\mathcal L_\mu\cap\sigma_x|$ is smaller than $2$, and then that it is larger than 0.
  \begin{itemize}
    \item To prove that $|\mathcal L_\mu\cap\sigma_x|<2$, we show that if $y,y'\in\mathcal L_\mu\cap\sigma_x$, then $\sigma_y\cap\sigma_{y'}\neq\emptyset$. Indeed, since $y\in\sigma_x$, writing $y'=x+\upsilon\in\sigma_x$, by translating by $\upsilon$, we find that $\sigma_{y'}\equiv\sigma_{x+\upsilon}\ni y+\upsilon\in\sigma_y$. Therefore, $|\mathcal L_\mu\cap\sigma_x|<2$.
    \item Finally, if $|\mathcal L_\mu\cap\sigma_x|=0$, then, since $\mathcal L_\mu$ is periodic, the density of $\mathcal L_\mu$ would be $<\rho_m$, which contradicts the fact that the $\mathcal L_i$ are related to each other by isometries.
  \end{itemize}
  All in all, $|\mathcal L_\mu\cap\sigma_x|=1$,which concludes the proof of~\-(\ref{assumK_0}).
  \bigskip

  \point From now on, we assume that~\-(\ref{assumK}) holds for every tiled strict subset of $\Lambda$ (note that $\hat\Gamma_{\gamma,j}$ is a tiled strict subset of $\Lambda$). We first prove~\-(\ref{cvcd}).
  \bigskip
  
  \subpoint By~\-(\ref{zeta}) and~\-(\ref{assumK}),
  \begin{equation}
    |\zeta_\nu^{(\underline z)}(\gamma)|\leqslant e^{2\cst c{cst:z}}\varsigma^{h_{\Gamma_\gamma}}\frac{|z|^{|X_\gamma|}}{|z|^{\rho_m|\Gamma_\gamma|}}e^{\chi|\Gamma_\gamma|}
    \label{bound_zeta_pre}
  \end{equation}
  in which $\chi$ is the coordination number of $\Lambda_\infty$ ($\chi$ appears because, for any set $A\subset\Lambda_\infty$, $|\partial A|\leqslant\chi|\partial(\Lambda_\infty\setminus A)|$). By definition~\-\ref{def:GFc}, in every configuration $X_\gamma$, every particle must be in contact with at least one empty site. Therefore, the fraction $\psi_\gamma(X_\gamma)$ of empty sites in $\Gamma_\gamma$ must satisfy
  \begin{equation}
    \psi_\gamma(X_\gamma):=\frac{|\mathcal E_{\Gamma_\gamma}(X_\gamma)|}{|\Gamma_\gamma|}\geqslant
    \frac1{\mathcal N+1}
  \end{equation}
  (recall that $|\mathcal E_{\Gamma_\gamma}(X_\gamma)|$ is the number of empty sites~\-(\ref{mcE}), and $\mathcal N$ is the maximal volume occupied by particles that neighbor a site~\-(\ref{mcN})). Therefore,
  \begin{equation}
    |X_\gamma|
    =
    \rho_m|\Gamma_\gamma|(1-\psi_\gamma(X_\gamma))\leqslant\rho_m|\Gamma_\gamma|\frac{\mathcal N}{\mathcal N+1}.
  \end{equation}
  Therefore, by~\-(\ref{alpha}), (\ref{bound_varsigma}) and~\-(\ref{bound_zeta_pre}), and using the fact that $h_{\Gamma_\gamma}\leqslant|\Gamma_\gamma|$,
  \begin{equation}
    |\zeta_\nu^{(\underline z)}(\gamma)|\leqslant
    \varsigma\left(\varsigma e^\chi|z|^{-\rho_m\frac1{\mathcal N+1}}\right)^{|\Gamma_\gamma|}
    \equiv\varsigma\alpha^{|\Gamma_\gamma|}
    .
    \label{bound_zeta}
  \end{equation}
  Thus, by~\-(\ref{a}),
  \begin{equation}
    |\zeta_\nu^{(\underline z)}(\gamma)|e^{a(\gamma)+d(\gamma)}\leqslant\varsigma\alpha^{(1-(\theta+\xi))|\Gamma_\gamma|}
    \label{bound_zetaead}
  \end{equation}
  which proves the first inequality in~\-(\ref{cvcd}) with $\delta\equiv\varsigma\alpha^{1-(\theta+\xi)}$, which, provided
  \begin{equation}
    \alpha\ll\varsigma^{-(1-(\theta+\xi))^{-1}}
    \label{assum_alpha1}
  \end{equation}
  satisfies $\delta\ll 1$.
  \bigskip

  \subpoint We now turn to the second inequality in~\-(\ref{cvcd}). By~\-(\ref{bound_zetaead}),
  \begin{equation}
    \sum_{\displaystyle\mathop{\scriptstyle\gamma'\in\mathfrak C_\nu(\Lambda)}_{\gamma'\not\sim\gamma}}e^{a(\gamma')+d(\gamma')}|\zeta_\nu^{(\underline z)}(\gamma')|
    \leqslant
    \varsigma
    \sum_{\displaystyle\mathop{\scriptstyle\gamma'\in\mathfrak C_\nu(\Lambda)}_{\gamma'\not\sim\gamma}}
    \alpha^{(1-(\theta+\xi))|\Gamma_{\gamma'}|}.
  \end{equation}
  We bound the number of GFcs $\gamma'$ that are {\it incompatible} with a fixed GFc $\gamma$ by the number of walks on $\Lambda_\infty$ of length $2|\Gamma_{\gamma'}|\equiv2\ell$ that intersect or neighbor $\Gamma_\gamma$:
  \begin{equation}
    \sum_{\displaystyle\mathop{\scriptstyle\gamma'\in\mathfrak C_\nu(\Lambda)}_{\gamma'\not\sim\gamma}}e^{a(\gamma')+d(\gamma')}|\zeta_\nu^{(\underline z)}(\gamma')|
    \leqslant
    \varsigma
    (\chi+1)|\Gamma_\gamma|\sum_{\ell=1}^\infty \chi^{2\ell}\alpha^{(1-(\theta+\xi))\ell}
  \end{equation}
  ($(\chi+1)|\Gamma_\gamma|$ is a bound on the number of sites that intersect or neighbor $\Gamma_\gamma$). Now, provided
  \begin{equation}
    \alpha\ll\chi^{-2(1-(\theta+\xi))^{-1}}
    \label{assum_alpha2}
  \end{equation}
  we have
  \begin{equation}
    \sum_{\displaystyle\mathop{\scriptstyle\gamma'\in\mathfrak C_\nu(\Lambda)}_{\gamma'\not\sim\gamma}}e^{a(\gamma')+d(\gamma')}|\zeta_\nu^{(\underline z)}(\gamma')|
    \leqslant
    \varsigma
    \cst c{cst:sumup}
    |\Gamma_\gamma|
    \label{bound_entropy1}
  \end{equation}
  for some constant $\cst c{cst:sumup}>0$. If, in addition,
  \begin{equation}
    \alpha\ll e^{-\varsigma\cst c{cst:sumup}\theta^{-1}}
    \label{assum_alpha3}
  \end{equation}
  then this implies~\-(\ref{cvcd}).
  \bigskip

  \point Let us now prove~\-(\ref{bound_deriv}).
  Since~\-(\ref{cvcd}) holds, the cluster expansion in lemma~\-\ref{lemma:cluster_expansion} is absolutely convergent. Thus, by~\-(\ref{ce}),
  \begin{equation}
    \frac\partial{\partial\log z(\mathfrak x_i)}\log\left(
      \frac{\Xi_\Lambda^{(\mu)}(\underline z)}{\mathbf z_\mu(\Lambda)}
    \right)
    =
    \sum_{\gamma'\in\mathfrak C_\mu(\Lambda)}
    \frac{\partial \zeta_\mu^{(\underline z)}(\gamma')}{\partial\log z(\mathfrak x_i)}
    \sum_{\underline\gamma\sqsubset\mathfrak C_\mu(\Lambda)}\Phi^T(\{\gamma'\}\sqcup\underline\gamma)\prod_{\gamma\in\underline\gamma}\zeta_\mu^{(\underline z)}(\gamma)
  \end{equation}
  so, by~\-(\ref{ce_remainder}),
  \begin{equation}
    \left|\frac\partial{\partial\log z(\mathfrak x_i)}\log\left(
      \frac{\Xi_\Lambda^{(\mu)}(\underline z)}{\mathbf z_\mu(\Lambda)}
    \right)\right|
    \leqslant
    \sum_{\gamma'\in\mathfrak C_\mu(\Lambda)}
    e^{a(\gamma')}
    \left|\frac{\partial \zeta_\mu^{(\underline z)}(\gamma')}{\partial\log z(\mathfrak x_i)}\right|.
    \label{bound_deriv1}
  \end{equation}
  Furthermore, by~\-(\ref{zeta}),
  \begin{equation}
    \begin{array}{>\displaystyle r@{\ }>\displaystyle l}
      \frac{\partial\log\zeta_\mu^{(\underline z)}(\gamma')}{\partial\log z(\mathfrak x_i)}
      =&
      \mathds 1\left(\mathfrak x_i\in X_{\gamma'}\right)
      -
      \mathds 1\left(\mathfrak x_i\in\mathcal L_\mu\cap\Gamma_{\gamma'}\right)
      \\[0.5cm]
      &+
      \sum_{j=1}^{h_{\Gamma_{\gamma'}}}\left(
	\mathds 1\left(\mathfrak x_i\in\mathcal L_{\underline\mu_{\gamma'}(\hat\Gamma_{\gamma',j})}\cap\hat\Gamma_{\gamma',j}\right)
	-
	\mathds 1\left(\mathfrak x_i\in\mathcal L_\mu\cap\hat\Gamma_{\gamma',j}\right)
      \right)
      \\[0.5cm]
      &+
      \sum_{j=1}^{h_{\Gamma_{\gamma'}}}\left(
	\frac\partial{\partial\log z(\mathfrak x_i)}\log\left(
	  \frac{\Xi_{\hat\Gamma_{\gamma',j}}^{(\underline\mu_{\gamma'}(\hat\Gamma_{\gamma',j}))}(\underline z)}{\mathbf z_{\underline\mu_{\gamma'}(\hat\Gamma_{\gamma',j})}(\hat\Gamma_{\gamma',j})}
	\right)
	-
	\frac\partial{\partial\log z(\mathfrak x_i)}\log\left(
	  \frac{\Xi_{\hat\Gamma_{\gamma',j}}^{(\mu)}(\underline z)}{\mathbf z_\mu(\hat\Gamma_{\gamma',j})}
	\right)
      \right)
      .
    \end{array}
    \label{first_deriv}
  \end{equation}
  Therefore, using~\-(\ref{bound_deriv}) inductively to estimate the last term,
  \begin{equation}
    \left|\frac{\partial\zeta_\mu^{(\underline z)}(\gamma')}{\partial\log z(\mathfrak x_i)}\right|
    \leqslant
    |\zeta_\mu^{(\underline z)}(\gamma')|
    3\mathds 1(\mathfrak x_i\in\mathrm{Int}(\Gamma_{\gamma'}))
  \end{equation}
  in which
  \begin{equation}
    \mathrm{Int}(\Gamma_{\gamma'}):=
    \Gamma_{\gamma'}\cup
    \left(
      \bigcup_{j=1}^{h_{\Gamma_{\gamma'}}}\hat\Gamma_{\gamma',j}
    \right)
  \end{equation}
  so that
  \begin{equation}
    \left|\frac\partial{\partial\log z(\mathfrak x_i)}\log\left(
      \frac{\Xi_\Lambda^{(\mu)}(\underline z)}{\mathbf z_\mu(\Lambda)}
    \right)\right|
    \leqslant
    3
    \sum_{\displaystyle\mathop{\scriptstyle\gamma'\in\mathfrak C_\mu(\Lambda)}_{\mathrm{Int}(\Gamma_{\gamma'})\ni\mathfrak x_i}}
    e^{a(\gamma')}
    |\zeta_\mu^{(\underline z)}(\gamma')|
    .
  \end{equation}
  In addition, by the isoperimetric inequality,
  \begin{equation}
    |\mathrm{Int}(\Gamma_{\gamma'})|
    \leqslant
    \cst c{cst:iso}^{(d)}|\Gamma_{\gamma'}|^d
  \end{equation}
  for some constant $\cst c{cst:iso}^{(d)}>0$ (which depends on $d$), so
  \begin{equation}
    \left|\frac\partial{\partial\log z(\mathfrak x_i)}\log\left(
      \frac{\Xi_\Lambda^{(\mu)}(\underline z)}{\mathbf z_\mu(\Lambda)}
    \right)\right|
    \leqslant
    3
    \sum_{\displaystyle\mathop{\scriptstyle\gamma'\in\mathfrak C_\mu(\Lambda)}_{\Gamma_{\gamma'}\ni\mathfrak x_i}}
    \cst c{cst:iso}^{(d)}|\Gamma_{\gamma'}|^d
    e^{a(\gamma')}
    |\zeta_\mu^{(\underline z)}(\gamma')|
    .
  \end{equation}
  Furthermore,
  \begin{equation}
    |\Gamma_{\gamma'}|^d\leqslant d! e^{|\Gamma_{\gamma'}|}
  \end{equation}
  so, rewriting
  \begin{equation}
     e^{a(\gamma')+|\Gamma_{\gamma'}|}=e^{-\bar d(\gamma')}e^{(a(\gamma')+d(\gamma'))}
     ,\quad
     \bar d(\gamma'):=d(\gamma)-|\Gamma_{\gamma'}|\geqslant-\xi\log\alpha-1
  \end{equation}
  which holds provided
  \begin{equation}
    \alpha\leqslant e^{-\frac1\xi}
    \label{assum_alpha4}
  \end{equation}
  and by~\-(\ref{bound_entropy1}), we find
  \begin{equation}
    \left|\frac\partial{\partial\log z(\mathfrak x_i)}\log\left(
      \frac{\Xi_\Lambda^{(\mu)}(\underline z)}{\mathbf z_\mu(\Lambda)}
    \right)\right|
    \leqslant
    \alpha^\xi 3e^1
    \cst c{cst:iso}^{(d)}
    d!
    \varsigma
    \cst c{cst:sumup}
    .
  \end{equation}
  which, provided
  \begin{equation}
    \alpha\leqslant \left(3e^1\cst c{cst:iso}^{(d)}d!\varsigma\cst c{cst:sumup}\right)^{-(\xi-\cst C{cst:deriv_Xi})^{-1}}
    \label{assum_alpha5}
  \end{equation}
  implies~\-(\ref{bound_deriv}).
  \bigskip

  \point We now turn to the proof of~\-(\ref{assumK}).
  \bigskip

  \subpoint First of all, we get rid of the dependence on $z(\mathfrak x_i)$: by Taylor's theorem,
  \begin{equation}
    \log\left(\frac{\Xi_{\Lambda}^{(\mu)}(\underline z)}{\Xi_{\Lambda}^{(\nu)}(\underline z)}\right)
    =
    \log\left(\frac{\Xi_{\Lambda}^{(\mu)}(z)}{\Xi_{\Lambda}^{(\nu)}(z)}\right)
    +
    \sum_{i=1}^{\mathfrak n}
    (\underline z(\mathfrak x_i)-z)
    \frac{\partial}{\partial\underline{\tilde z}(\mathfrak x_i)}
    \log\left(\frac{\Xi_{\Lambda}^{(\mu)}(\tilde{\underline z})}{\Xi_{\Lambda}^{(\nu)}(\tilde{\underline z})}\right)
    \label{taylor}
  \end{equation}
  in which $\tilde{\underline z}$ is a function satisfying $\tilde{\underline z}(\mathfrak x_i)\in[z,\underline z(\mathfrak x_i)]$ and $\tilde{\underline z}(x)=z$ for any $x\neq\mathfrak x_i$. By~\-(\ref{bound_deriv}),
  \begin{equation}
    \left|
      \frac{\partial}{\partial\underline{\tilde z}(\mathfrak x_i)}
      \log\left(\frac{\Xi_{\Lambda}^{(\mu)}(\tilde{\underline z})}{\Xi_{\Lambda}^{(\nu)}(\tilde{\underline z})}\right)
    \right|
    \leqslant
    \frac1{|\tilde{\underline z}(\mathfrak x_i)|}
    \left(
      \left|
	\mathds 1\left(\mathfrak x_i\in\mathcal L_\mu\cap\Lambda\right)
	-
	\mathds 1\left(\mathfrak x_i\in\mathcal L_\nu\cap\Lambda\right)
      \right|
      +
      \alpha^{\cst C{cst:deriv_Xi}}
    \right)
    .
  \end{equation}
  Thus,
  \begin{equation}
    \left|
      \sum_{i=1}^{\mathfrak n}
      (\underline z(\mathfrak x_i)-z)
      \frac{\partial}{\partial\underline{\tilde z}(\mathfrak x_i)}
      \log\left(\frac{\Xi_{\Lambda}^{(\mu)}(\tilde{\underline z})}{\Xi_{\Lambda}^{(\nu)}(\tilde{\underline z})}\right)
    \right|
    \leqslant
    2\mathfrak n(e^{\frac{2\cst c{cst:z}}{\mathfrak n}}+1)
    .
    \label{taylor_remainder}
  \end{equation}
  \bigskip

  \subpoint We now focus on $\Xi_\Lambda^{(\mu)}(z)$, and make use of the cluster expansion in lemma~\-\ref{lemma:cluster_expansion}: by~\-(\ref{ce}),
  \begin{equation}
    \log\left(\frac{\Xi_{\Lambda}^{(\mu)}(z)}{\Xi_{\Lambda}^{(\nu)}(z)}\right)
    =
    \sum_{\underline\gamma\sqsubset\mathfrak C_\mu(\Lambda)}\Phi^T(\underline\gamma)\prod_{\gamma\in\underline\gamma}\zeta_\mu^{(z)}(\gamma)
    -
    \sum_{\underline\gamma\sqsubset\mathfrak C_\nu(\Lambda)}\Phi^T(\underline\gamma)\prod_{\gamma\in\underline\gamma}\zeta_\nu^{(z)}(\gamma)
    \label{flip_ce}
  \end{equation}
  (we recall that $z^{|\Lambda\cap\mathcal L_\mu|}$ is independent of $\mu$ so the $\mathbf z_\mu(\Lambda)$ and $\mathbf z_\nu(\Lambda)$ factors cancel out). We then split these cluster expansions into {\it bulk} and {\it boundary} contributions, which are defined as follows. Let $\mathfrak C_\mu^{(|\Lambda|)}(\Lambda_\infty)$ denote the set of GFcs in $\Lambda_\infty$ whose upper-leftmost corner (if $d>2$, then this notion should be extended in the obvious way) is in $\Lambda$. Note that $\mathfrak C_\mu^{(|\Lambda|)}(\Lambda_\infty)$ only depends on $\Lambda$ through its cardinality $|\Lambda|$ (up to a translation). We then write
  \begin{equation}
    \sum_{\underline\gamma\subset\mathfrak C_\mu(\Lambda)}\Phi^T(\underline\gamma)\prod_{\gamma\in\underline\gamma}\zeta_\mu^{(z)}(\gamma)
    =
    \mathfrak B^{(|\Lambda|)}_\mu(\Lambda_\infty)
    -
    \mathfrak b_\mu^{(\Lambda)}(\Lambda_\infty)
    \label{bulk_boundary}
  \end{equation}
  in which $\mathfrak B$ is the {\it bulk} contribution, and $\mathfrak b$ is the {\it boundary} term.
  \begin{equation}
    \begin{array}{>\displaystyle l}
      \mathfrak B^{(|\Lambda|)}_\mu(\Lambda_\infty):=
      \sum_{m=1}^\infty\sum_{\gamma'\in\mathfrak C_\mu^{(|\Lambda|)}(\Lambda_\infty)}
      (\zeta_\mu^{(z)}(\gamma'))^m
      \sum_{\underline\gamma\sqsubset\mathfrak C_\mu(\Lambda_\infty)\setminus\{\gamma'\}}
      \Phi^T(\{\gamma'\}^m\sqcup\underline\gamma)\prod_{\gamma\in\underline\gamma}\zeta_\mu^{(z)}(\gamma)
      \\[1cm]
      \mathfrak b^{(\Lambda)}_\mu(\Lambda_\infty):=
      \sum_{m=1}^\infty\sum_{\gamma'\in\mathfrak C_\mu^{(|\Lambda|)}(\Lambda_\infty)}
      (\zeta_\mu^{(z)}(\gamma'))^m
      \sum_{\displaystyle\mathop{\scriptstyle\underline\gamma\sqsubset\mathfrak C_\mu(\Lambda_\infty)\setminus\{\gamma'\}}_{(\{\gamma'\}^m\sqcup\underline\gamma)\not\sqsubset\mathfrak C_\mu(\Lambda)}}
      \Phi^T(\{\gamma'\}^m\sqcup\underline\gamma)\prod_{\gamma\in\underline\gamma}\zeta_\mu^{(z)}(\gamma)
    \end{array}
    \label{frakB}
  \end{equation}
  in which $\{\gamma'\}^m$ is the multiset with $m$ elements that are all equal to $\gamma'$.
  \bigskip

  \subsubpoint The bulk terms cancel each other out. Indeed, we recall (see section~\-\ref{sec:model}) that there exists an isometry $F_{\mu,\nu}$ of $\Lambda_\infty$ such that $F_{\mu,\nu}(\mathcal L_\mu)=\mathcal L_\nu$. In addition, since $F_{\mu,\nu}$ is an isometry, it maps perfect coverings to perfect coverings, and this map is denoted by $f_{\mu,\nu}:\{1,\cdots,\tau\}\to\{1,\cdots,\tau\}$:
  \begin{equation}
    \mathcal L_{f_{\mu,\nu}(\kappa)}=F_{\mu,\nu}(\mathcal L_\kappa).
  \end{equation}
  This allows us to define an action on GFcs: $\mathfrak F_{\mu,\nu}:\mathfrak C_\mu(\Lambda)\to\mathfrak C_\nu(F_{\mu,\nu}(\Lambda))$,
  \begin{equation}
    \mathfrak F_{\mu,\nu}(\Gamma_\gamma,X_\gamma,\mu,\underline\mu_\gamma):=(F_{\mu,\nu}(\Gamma_\gamma),F_{\mu,\nu}(X_\gamma),\nu,f_{\mu,\nu}(\underline\mu_\gamma))
    .
  \end{equation}
  The map $\mathfrak F_{\mu,\nu}$ is a bijection and, since the partition function is invariant under isometries, it leaves $\zeta_\mu^{(z)}$ and $\Phi^T$ invariant, so
  \begin{equation}
    \mathfrak B_\mu^{(|\Lambda|)}(\Lambda_\infty)
    =
    \sum_{m=1}^\infty\sum_{\gamma'\in\mathfrak C_\nu^{(|F_{\mu,\nu}(\Lambda)|)}(F_{\mu,\nu}(\Lambda_\infty))}
    (\zeta_\nu^{(z)}(\gamma'))^m
    \sum_{\underline\gamma\sqsubset\mathfrak C_\nu(F_{\mu,\nu}(\Lambda_\infty))\setminus\{\gamma'\}}
    \Phi^T(\{\gamma'\}^m\sqcup\underline\gamma)\prod_{\gamma\in\underline\gamma}\zeta_\nu^{(z)}(\gamma)
    \label{frakBrot}
  \end{equation}
  so, since $F_{\mu,\nu}(\Lambda_\infty)=\Lambda_\infty$ and $|F_{\mu,\nu}(\Lambda)|=|\Lambda|$,
  \begin{equation}
    \mathfrak B_\mu^{(|\Lambda|)}(\Lambda_\infty)
    -
    \mathfrak B_\nu^{(|\Lambda|)}(\Lambda_\infty)
    =0
    .
    \label{bulk}
  \end{equation}
  \bigskip

  \subsubpoint Finally, we estimate the boundary term. First of all, since every cluster $\{\gamma'\}\sqcup\underline\gamma$ that is not a subset of $\mathfrak C_\mu(\Lambda)$ must contain at least one GFc that goes over the boundary of $\Lambda$,
  \begin{equation}
    \mathfrak b_\mu^{(\Lambda)}(\Lambda_\infty)
    \leqslant
    \sum_{\displaystyle\mathop{\scriptstyle\gamma'\in\mathfrak C_\nu(\Lambda_\infty)}_{\displaystyle\mathop{\scriptstyle\Gamma_{\gamma'}\cap\Lambda\neq\emptyset}_{\Gamma_{\gamma'}\cap(\Lambda_\infty\setminus\Lambda)\neq\emptyset}}}
    |\zeta_\mu^{(z)}(\gamma')|
    \sum_{\underline\gamma\sqsubset\mathfrak C_\mu(\Lambda_\infty)}
    \left|\Phi^T(\{\gamma'\}\sqcup\underline\gamma)\prod_{\gamma\in\underline\gamma}\zeta_\mu^{(z)}(\gamma)\right|
  \end{equation}
  (for the purpose of an upper bound, we can reabsorb the sum over $m$ in~\-(\ref{frakB}) in the sum over $\underline\gamma$) so, by~\-(\ref{ce_remainder}),
  \begin{equation}
    |\mathfrak b_\mu^{(\Lambda)}(\Lambda_\infty)|
    \leqslant
    \sum_{\displaystyle\mathop{\scriptstyle\gamma\in\mathfrak C_\nu(\Lambda_\infty)}_{\displaystyle\mathop{\scriptstyle\Gamma_{\gamma}\cap\Lambda\neq\emptyset}_{\Gamma_{\gamma}\cap(\Lambda_\infty\setminus\Lambda)\neq\emptyset}}}
    |\zeta_\mu^{(z)}(\gamma')|
    e^{a(\gamma')}
  \end{equation}
  which, rewriting, as we did earlier $e^{a(\gamma')}=e^{-d(\gamma')}e^{a(\gamma')+d(\gamma')}$ and using $d(\gamma')\geqslant-\xi\log\alpha$, implies, similarly to the derivation of~\-(\ref{bound_entropy1}),
  \begin{equation}
    |\mathfrak b_\mu^{(\Lambda)}(\Lambda_\infty)|
    \leqslant
    \alpha^\xi\varsigma\cst c{cst:sumup}|\partial\Lambda|.
    \label{boundary}
  \end{equation}
  \bigskip

  \subsubpoint Thus, inserting~\-(\ref{bulk}) and~\-(\ref{boundary}) into~\-(\ref{bulk_boundary}) and~\-(\ref{flip_ce}), provided
  \begin{equation}
    2\alpha^\xi\varsigma\cst c{cst:sumup}\leqslant 1
    \label{assum_alpha6}
  \end{equation}
  we find that
  \begin{equation}
    \log\left(\frac{\Xi_{\Lambda}^{(\mu)}(z)}{\Xi_{\Lambda}^{(\nu)}(z)}\right)
    \leqslant
    |\partial\Lambda|.
  \end{equation}
  By combining this bound with~\-(\ref{taylor_remainder}) and~\-(\ref{taylor}), we find that~\-(\ref{assumK}) holds with
  \begin{equation}
    \varsigma=1+2\mathfrak n(e^{2\frac{\cst c{cst:z}}{\mathfrak n}}+1).
    \label{varsigma}
  \end{equation}
\qed

\subsection{High-fugacity expansion}
\indent We now conclude this section by summarizing the validity of the high-fugacity expansion as a stand-alone theorem, which is a simple consequence of lemmas~\-\ref{lemma:GFc}, \ref{lemma:cluster_expansion} and~\-\ref{lemma:bound_zeta}, and showing how it implies theorem~\-\ref{theo:main}.
\bigskip

\theoname{Theorem}{high-fugacity expansion}\label{theo:expansion}
  Consider a non-sliding hard-core lattice particle system and a boundary condition $\nu\in\{1,\cdots,\tau\}$. We assume that $z(x)$ takes the same value $z$ for every $x\in\Lambda_\infty$ except for a finite number $\mathfrak n$ of sites $(\mathfrak x_1,\cdots,\mathfrak x_{\mathfrak n})$ (that is, $z(x)=z$ for every $x\in\Lambda_\infty\setminus\{\mathfrak x_1,\cdots,\mathfrak x_{\mathfrak n}\}$). There exists $z_0,\cst c{cst:z}>0$ such that if
  \begin{equation}
    |z|>z_0
    ,\quad
    e^{-\frac{\cst c{cst:z}}{\mathfrak n}}|z|
    \leqslant
    |z(\mathfrak x_i)|
    \leqslant
    e^{\frac{\cst c{cst:z}}{\mathfrak n}}|z|
  \end{equation}
  then the following hold.
  \bigskip

  The partition function~\-(\ref{Xi_nu}) can be rewritten as
  \begin{equation}
    \frac{\Xi_\Lambda^{(\nu)}(\underline z)}{\mathbf z_\nu(\Lambda)}=\exp\left(\sum_{\underline\gamma\sqsubset\mathfrak C_\nu(\Lambda)}\Phi^T(\underline\gamma)\prod_{\gamma\in\underline\gamma}\zeta_\nu^{(\underline z)}(\gamma)\right)
    \label{ce_theo}
  \end{equation}
  where $\mathbf z_\nu(\Lambda)$ and $\zeta_\nu^{(\underline z)}(\gamma)$ were defined in~\-(\ref{bfz}) and~\-(\ref{zeta}), and $\Phi^T$ was defined in~\-(\ref{PhiT}).
  \bigskip

  In addition, (\ref{ce_theo}) is absolutely convergent: there exist $\epsilon,\cst C{cst:cvce}>0$, such that, for every $\gamma'\in\mathfrak C_\nu(\Lambda)$,
  \begin{equation}
    \sum_{\underline\gamma\sqsubset\mathfrak C_\nu(\Lambda)}
    \left|
      \Phi^T(\{\gamma'\}\sqcup\underline\gamma)
      \zeta_\nu^{(\underline z)}(\gamma')\prod_{\gamma''\in\underline\gamma}\zeta_\nu^{(\underline z)}(\gamma'')
    \right|
    \leqslant
    \cst C{cst:cvce}\epsilon^{|\Gamma_\gamma|}
    \label{ineq_ce}
  \end{equation}
  and $\epsilon\to0$ as $y\equiv z^{-1}\to0$.
\endtheo
\bigskip

{\bf Remark}: The quantities $z_0$, $\epsilon$ and $\cst C{cst:cvce}$ depend on the model. They are computed above (see lemma~\-\ref{lemma:bound_zeta}), although we do not expect that the expressions given in this paper are anywhere near optimal. Instead, the take-home message we would like to convey here, is that these constants exist, and that $\epsilon$ is arbitrarily small (at the price of making the activity larger).
\bigskip

\indent Theorem~\-\ref{theo:main} is a corollary of theorem~\-\ref{theo:expansion}, as detailed below.
\bigskip

{\bf Proof of theorem~\-\ref{theo:main}}:\par\penalty10000\medskip\penalty10000
  \point By~\-(\ref{ce_theo}), the finite volume pressure is given by
  \begin{equation}
    p_\Lambda^{(\nu)}(z)=\frac1{|\Lambda|}\log\Xi_\Lambda^{(\nu)}=\frac1{|\Lambda|}\log\mathbf z_\nu(\Lambda)+
    \frac1{|\Lambda|}
    \sum_{\underline\gamma\sqsubset\mathfrak C_\nu(\Lambda)}\Phi^T(\underline\gamma)\prod_{\gamma\in\underline\gamma}\zeta_\nu^{(z)}(\gamma).
  \end{equation}
  Furthermore,
  \begin{equation}
    \log\mathbf z_\nu(\Lambda)=\rho_m|\Lambda|\log z
    .
  \end{equation}
  Now, by~\-(\ref{zeta}), $\zeta_\nu^{(z)}(\gamma)$ is a rational function of $y$, and, by~\-(\ref{cvcd}), it is bounded by 1 for small $y$, uniformly in $\gamma$. It is, therefore, an analytic function of $y$ for small $y$. In addition, $p_\Lambda^{(\nu)}(z)$ converges in the $\Lambda\to\Lambda_\infty$ limit uniformly in $y$, indeed, splitting into bulk and boundary terms as in~\-(\ref{bulk_boundary}), we find that the bulk term $\frac1{|\Lambda|}\mathfrak B_\nu^{(|\Lambda|)}(\Lambda_\infty)$ is independent of $\Lambda$, and that the boundary term $\frac1{|\Lambda|}\mathfrak b_\nu^{(\Lambda)}(\Lambda_\infty)$ vanishes in the infinite-volume limit~\-(\ref{boundary}). Therefore,
  \begin{equation}
    p(z)=
    \rho_m\log z
    +
    \frac1{|\Lambda|}
    \mathfrak B_\nu^{(|\Lambda|)}(\Lambda_\infty)
    .
  \end{equation}
  Furthermore, by lemma~\-\ref{lemma:cluster_expansion}, the sums over $\gamma'$ and $\underline\gamma$ in $\frac1{|\Lambda|}\mathfrak B_\nu^{(|\Lambda|)}(\Lambda_\infty)$ (see~\-(\ref{frakB})) are absolutely convergent, which implies that $p(z)-\rho_m\log z$ is an analytic function of $y$ for small value of $|y|$.
  \bigskip

  \point By a similar argument, we show that the correlation functions are analytic in $y$ for smallvalues of $|y|$ by proving that
  \begin{equation}
    \sum_{\underline\gamma\sqsubset\mathfrak C_\nu(\Lambda)}
    \frac{\partial^{\mathfrak n}}{\partial\log\underline z(\mathfrak x_1)\cdots\partial\log\underline z(\mathfrak x_{\mathfrak n})}
    \Phi^T(\underline\gamma)\prod_{\gamma\in\underline\gamma}
    \zeta_\nu^{(\underline z)}(\gamma)
  \end{equation}
  converges to
  \begin{equation}
    \sum_{\underline\gamma\sqsubset\mathfrak C_\nu(\Lambda_\infty)}
    \frac{\partial^{\mathfrak n}}{\partial\log\underline z(\mathfrak x_1)\cdots\partial\log\underline z(\mathfrak x_{\mathfrak n})}
    \Phi^T(\underline\gamma)\prod_{\gamma\in\underline\gamma}
    \zeta_\nu^{(\underline z)}(\gamma)
  \end{equation}
  uniformly in $y$, or, in other words, that their difference
  \begin{equation}
    \sum_{m=1}^\infty\sum_{\gamma'\in\mathfrak C_\nu(\Lambda_\infty)\setminus\mathfrak C_\nu(\Lambda)}
    \sum_{\underline\gamma\sqsubset\mathfrak C_\nu(\Lambda_\infty)\setminus\{\gamma'\}}
    \frac{\partial^{\mathfrak n}}{\partial\log\underline z(\mathfrak x_1)\cdots\partial\log\underline z(\mathfrak x_{\mathfrak n})}
    \Phi^T(\{\gamma'\}^m\sqcup\underline\gamma)
    (\zeta_\nu^{(\underline z)}(\gamma'))^m
    \prod_{\gamma\in\underline\gamma}\zeta_\nu^{(\underline z)}(\gamma)
    \label{remainder_correlations}
  \end{equation}
  vanishes in the infinite-volume limit. It is straightforward to check (this is done in detail for the first derivative in the proof of lemma~\-\ref{lemma:bound_zeta}, see~\-(\ref{first_deriv})) that the derivatives of $\log\zeta_\nu^{(\underline z)}(\gamma)$ are bounded analytic functions of $y$, uniformly in $\gamma$, and are proportional to indicator functions that force $\Gamma_\gamma$ to contain each of the $\mathfrak x_i$ with respect to which $\zeta$ is derived. Therefore, the clusters $\{\gamma'\}\sqcup\underline\gamma$ that contribute are those which contain all the $\mathfrak x_i$ and that are not contained inside $\Lambda$. We can therefore bound~\-(\ref{remainder_correlations}) by
  \begin{equation}
    \sum_{\displaystyle\mathop{\scriptstyle\gamma'\in\mathfrak C_\nu(\Lambda_\infty)}_{\Gamma_{\gamma'}\ni\mathfrak x_1}}
    \sum_{\underline\gamma\sqsubset\mathfrak C_\nu(\Lambda_\infty)}
    \left|
      \Phi^T(\{\gamma'\}\sqcup\underline\gamma)
      \zeta_\nu^{(\underline z)}(\gamma')
      \prod_{\displaystyle\mathop{\scriptstyle\gamma\in\underline\gamma}_{\mathrm{vol}(\{\gamma'\}\sqcup\underline\gamma)\geqslant\mathrm{dist}(\mathfrak x_1,\Lambda_\infty\setminus\Lambda)}}
      \zeta_\nu^{(\underline z)}(\gamma)
    \right|
  \end{equation}
  in which $\mathrm{vol}(\{\gamma'\}\sqcup\underline\gamma):=|\Gamma_{\gamma'}|+\sum_{\gamma\in\underline\gamma}|\Gamma_\gamma|$. By proceeding as in~\-(\ref{boundary}), we bound this contribution by
  \begin{equation}
    \cst c{cst:corr}
    \alpha^{\xi\mathrm{dist}(\mathfrak x_1,\Lambda_\infty\setminus\Lambda)}
  \end{equation}
  for some constant $\cst c{cst:corr}>0$, so it vanishes as $\Lambda\to\Lambda_\infty$. Furthermore, by the same argument, we show that the sum over $\underline\gamma$ in
  \begin{equation}
    \frac{\partial^{\mathfrak n}}{\partial\log\underline z(\mathfrak x_1)\cdots\partial\log\underline z(\mathfrak x_{\mathfrak n})}
    \sum_{\underline\gamma\sqsubset\mathfrak C_\nu(\Lambda_\infty)}\Phi^T(\underline\gamma)\prod_{\gamma\in\underline\gamma}\zeta_\nu^{(z)}(\gamma)
  \end{equation}
  is absolutely convergent, so
  \begin{equation}
    \frac{\partial^{\mathfrak n}}{\partial\log\underline z(\mathfrak x_1)\cdots\partial\log\underline z(\mathfrak x_{\mathfrak n})}
    \sum_{\underline\gamma\sqsubset\mathfrak C_\nu(\Lambda)}\Phi^T(\underline\gamma)\prod_{\gamma\in\underline\gamma}\zeta_\nu^{(z)}(\gamma)
  \end{equation}
  is analytic in $y$ for small $|y|$. Finally,
  \begin{equation}
    \frac{\partial^{\mathfrak n}}{\partial\log\underline z(\mathfrak x_1)\cdots\partial\log\underline z(\mathfrak x_{\mathfrak n})}
    \log\mathbf z_\nu(\Lambda)=
    \mathds 1(\mathfrak n=1)
    \mathds 1(\mathfrak x_1\in\mathcal L_\nu\cap\Lambda)
  \end{equation}
  which is, obviously, analytic in $y$. Therefore, the $\mathfrak n$-point truncated correlation functions are analytic in $y$ as well.
  \bigskip

  \point In particular, $\rho_1^{(\nu)}(x)$ is an analytic function of $y$, and its 0-th order term is the indicator function that $x\in\mathcal L_\nu$, which proves~\-(\ref{crystallization}). Finally $\rho_m-\rho$ is an analytic function of $y$,
  \begin{equation}
    \rho_m-\rho=c_1y+O(y^2)
    ,\quad
    c_1=\lim_{\Lambda\to\Lambda_\infty}\frac1{|\Lambda|}Q_\Lambda(1)\geqslant 1
  \end{equation}
  (we recall that $Q_\Lambda(1)$ is the number of particle configurations with $N_{\mathrm{max}}-1$ particles, which is at least $|\Lambda|$). Therefore $y\mapsto\rho_m-\rho$ is invertible, so the correlation functions and $p-\log(z)$ are also analytic functions of $\rho_m-\rho$. In addition, $\log(z)+\log(\rho_m-\rho)$ is analytic in $\rho_m-\rho$ as well.
\qed

\vfill
\hfil{\bf Acknowledgements}\par
\medskip
We are grateful to Giovanni Gallavotti and Roman Koteck\'y for enlightening discussions. The work of J.L.L. was supported by AFOSR grant FA9550-16-1-0037. The work of I.J. was supported by The Giorgio and Elena Petronio Fellowship Fund and The Giorgio and Elena Petronio Fellowship Fund II.
\vfil
\eject

\end{document}